\documentclass[twocolumn]{aastex63}

\usepackage{array}
\usepackage{xcolor}
\newcommand{\kep}       {{\it Kepler}}

\shorttitle{TEOs in Heartbeat Stars}
\shortauthors{Cheng et al.}

\graphicspath{{./}{figures/}}

\begin{document}

\title{Detailed Characterization of Heartbeat Stars and their Tidally Excited Oscillations}


\correspondingauthor{Shelley J. Cheng}
\email{shelleycheng@ucla.edu}

\author{Shelley J. Cheng}
\affiliation{Department of Physics and Astronomy, University of California, Los Angeles, CA 90095, USA}
\affiliation{TAPIR, Walter Burke Institute for Theoretical Physics, Mailcode 350-17, California Institute of Technology, Pasadena, CA 91125, USA}

\author{Jim Fuller}
\affiliation{TAPIR, Walter Burke Institute for Theoretical Physics, Mailcode 350-17, California Institute of Technology, Pasadena, CA 91125, USA}

\author{Zhao Guo}
\affiliation{Center for Exoplanets and Habitable Worlds, Department of Astronomy \& Astrophysics, 525 Davey Laboratory, The Pennsylvania State University, University Park, PA 16802, USA}

\author{Holger Lehman}
\affiliation{Th\"uringer Landessternwarte Tautenburg, Sternwarte 5, 07778 Tautenburg, Germany}

\author{Kelly Hambleton}
\affiliation{Department of Astrophysics and Planetary Science, Villanova University, 800 East Lancaster Avenue, Villanova, PA 19085, USA}

\begin{abstract}
Heartbeat stars are a class of eccentric binary stars with short-period orbits and characteristic ``heartbeat'' signals in their light curves at periastron, 
caused primarily by tidal distortion.
In many heartbeat stars, tidally excited oscillations can be observed throughout the orbit, with frequencies at exact integer multiples of the orbital frequency.
Here, we characterize the tidally excited oscillations in the heartbeat stars KIC 6117415, KIC 11494130, and KIC 5790807. Using \kep\ light curves and radial velocity measurements, we first model the heartbeat stars using the binary modeling software ELLC, including gravity darkening, limb darkening, Doppler boosting, and reflection. We then conduct a frequency analysis to determine the amplitudes and frequencies of the tidally excited oscillations.
Finally, we apply tidal theories to stellar structure models of each system to determine whether chance resonances can be responsible for the observed tidally excited oscillations, or whether a resonance locking process is at work. We find that resonance locking is likely occurring in KIC 11494130, but not in KIC 6117415 or KIC 5790807.
\end{abstract}


\section{Introduction} \label{sec:intro}
Heartbeat stars are binary stars with eccentric orbits that experience strong tidal interactions at periastron. These tidal interactions alter the observed cross sections and surface temperatures of both components.
Thus, the light curves feature a distinct ``heartbeat" signal near periastron, whose morphology is determined mostly by eccentricity and viewing angle.

Prior to \textit{Kepler}, only a small number of heartbeat stars were known, most notably HD 209295 \citep{Handler+02} and HD 174884 \citep{Maceroni+09}. However, with \textit{Kepler}, a large number of heartbeat stars have been discovered and analyzed, such as KOI-54 \citep{Welsh+11}, KIC 4544587 \citep{Hambleton+13}, KIC 10080943 \citep{Schmid+15}, KIC 3749404 \citep{Hambleton+16}, KIC 3230227 \citep{Guo+17}, KIC 4142768 \citep{Guo+19}, $15$ sub-giant/red giant heartbeat stars by \citet{Beck+14}, and a few others in \cite{Guo+20}. Several O- and B-type massive heartbeat stars have been discovered by BRITE and TESS, such as $\iota$ Orionis \citep{Pablo+17}, $\epsilon$ Lupi A \citep{Pablo+19}, and MACHO 80.7443.1718 \citep{Jayasinghe+19}. Additional spectral follow-up and characterization has been performed for a couple dozen other systems \citep{Smullen+15,Shporer+16,Dimitrov+17}. Thus far, 172 heartbeat star systems \citep{Kirk+16} have been catalogued in the \textit{Kepler} field, most of which have not been closely studied. 

Heartbeat stars have been observed to display a diverse range of characteristics, including tidally excited oscillations (TEOs). TEOs are driven by the varying tidal forces throughout the stars' orbit, and are hypothesized to play a significant role in the circularization of the binary's orbit \citep{Zahn+75, Goldreich+89, Witte+01}. While all heartbeat stars likely contain TEOs at some level, empirically only $\sim \! 20\%$ of the current \textit{Kepler} heartbeat star sample show clear TEOs \citep{Kirk+16}. These TEOs occur at exact integer multiples of the orbital frequency, with amplitudes dependent on the difference between the tidal forcing frequency and the tidally excited oscillation mode's natural frequency \citep{Kumar+95}, in addition to the properties of the star's oscillation modes. Since a near exact resonance between the tidal forcing frequency and a mode frequency is unlikely, high-amplitude TEOs are unlikely. Nonetheless, there appear to be several heartbeat star systems exhibiting high-amplitude TEOs potentially produced by near-resonances.

Resonance locking \citep{Witte+99, Witte+01} has been proposed as a mechanism that may explain the observation of high amplitude TEOs in heartbeat stars \citep{Fuller+12a,  Burkart+12, Burkart+14}, and has successfully explained the high-amplitude TEO in KIC 8164262 \citep{Hambleton+18,Fuller&Hambleton+17}. In resonance locking, a stable resonance between tidal forcing and a stellar oscillation mode is maintained by feedback between stellar and tidal evolution. Over time, stellar oscillation mode frequencies change due to factors such as stellar evolution, tidal spin-up, magnetic braking, etc. Likewise, orbital decay occurs due to tidal dissipation, changing the tidal forcing frequencies. If a system is near resonance, there exists a stable equilibrium in which the tidally excited oscillation causes the orbit to decay at precisely the rate required to maintain the resonance. This can result in a prominent and stable TEO. 

In this paper, we use photometric and radial velocity data to characterize the heartbeat star systems KIC 6117415, KIC 11494130, and KIC 5790807. Using existing data (\kep\ lightcurves and ground-based spectra - see Section~\ref{sec:data}), we develop binary models (Section~\ref{sec:model}) for each system and perform a frequency analysis (Section~\ref{sec:freq}). We then measure TEO amplitudes and model the TEOs based on the stellar and orbital properties (Section~\ref{sec:TEO}) following \citet{Fuller+17}. Finally, we investigate whether resonance locking can explain the observed TEOs (Section~\ref{sec:reslock}). A discussion of our results is offered in Section~\ref{sec:discuss} before drawing our conclusions.

\section{Observations} \label{sec:data}

Our chosen heartbeat stars are part of the \kep\ Eclipsing Binary Catalog. We selected KIC 6117415 for further study since it is a double lined eclipsing spectroscopic binary, and selected KIC 11494130 and KIC 5790807 since they visibly exhibit TEOs. For our modeling, we combined light curve data from \textit{Kepler} and radial velocity data from the HIRES spectrograph at the Keck telescope, the Th\"uringer Landessternwarte Tautenburg (TLS) Observatory, from \citet{Smullen+15}, and from \citet{Shporer+16}.

\subsection{\textit{Kepler} light curve data}
We used all available quarters of long cadence observations from the \kep\ telescope through the Mikulski Archive for Space Telescopes database, and created light curves using the barycentric times and \texttt{PDCSAP\textunderscore{}FLUX} fluxes reported in the \kep\ data. The \texttt{PDCSAP\textunderscore{}FLUX} fluxes were processed through the Presearch Data Conditioning (PDC) module of the \kep\ pipeline, which removes systematic trends and other instrumental signatures through a Bayesian maximum a posteriori approach \citep{Smith+12, Stumpe+12, Stumpe+14}. We then performed detrending using a Savitzky-Golay filter with a third order polynomial, and then folded the light curve using the reported \kep\ period.

Since the large number of data points in the folded \kep\ light curve is computationally expensive to model, we phase binned the light curve by taking the median of a specified number of data points in the phased light curve. Such binning is appropriate for our chosen heartbeat star systems since the TEOs and periastron features cyclically repeat. For KIC 6117415, which has a \textit{Kepler} amplitude of $\text{Kp} = 10.543$, we binned the light curve by taking the median of every $15$ data points. Due to the smaller number of data points in the eclipse in KIC 5790807, we binned every $3$ data points. Conversely, the lack of an eclipse in KIC 11494130 allowed us to bin every $40$ datapoints. 

\subsection{Keck/HIRES spectroscopy data}\label{sec:keck}

We gathered spectra of KIC 6117415 using Keck/HIRES as described in \citet{Shporer+16}. KIC 6117415 was not included as part of that publication because it is a double-lined system requiring different spectral disentangling and radial velocity extraction techniques. We measured the radial velocities by using the 2D cross-correlation technique based on \citet{Gies+86}. The observed spectra are compared with a composite template generated from the \texttt{BLUERED} library \citep{Bertone+08}. The absolute Doppler shift of the primary star and the relative radial velocity of the secondary are determined sequentially. The final radial velocities are listed in Table~\ref{table:6117rvs}. We then use the spectral separation algorithm \citep{Bagnuolo+94} to obtain the individual spectrum of the two components. The atmospheric parameters are determined by comparing the two individual spectra with synthetic spectra from the \texttt{BLUERED} library. The optimal values and their uncertainties are obtained by using the nested sampling package \texttt{MultiNest} \citet{Feroz+09} and are shown in Table \ref{table:6117mcmc}.

\subsection{TLS spectroscopy data} \label{sec:TLS}

We obtained 11 spectra of KIC\,11494130 with the TCES spectrograph\footnote{http://www.tls-tautenburg.de/TLS/index.php?id=31\&L=1} attached to the 2-m Alfred-Jensch-Telescope of the Th\"uringer Landessternware Tautenburg in 2012. The obtained spectra have a resolving power of 62\,000 and cover the wavelength range 4720 to 7350\,\AA.  Spectrum reduction included the filtering of cosmic ray events, flat fielding, wavelength calibration using a ThAr lamp, optimum extraction of the spectra, normalization to the local continuum, and the merging of the Echelle orders. We primarily used standard ESO-MIDAS routines, and our own routines for spectrum normalization and for correcting for nightly instrumental shifts in radial velocity (based on a large number of telluric O$_2$ lines).

Radial velocities were determined from cross-correlating the observed spectra with a synthetic template spectrum, calculated with SynthV \citep{1996ASPC..108..198T} based on atmosphere models computed with LLmodels \citep{2004A&A...428..993S} for $T_{\rm eff}$ = 6700\,K, $\log{g}$ = 4.5, $v\sin{i}$ = 1\,km\,s$^{-1}$, and $\lambda\lambda$\,=\,4915...5670\,\AA, assuming solar abundances. Atomic data were taken from the VALD data base \citep{2000BaltA...9..590K}. $T_{\rm eff}$ was chosen to be close to the value given in \citet{2012ApJ...753...86T}, the wavelength range was selected to exclude the broad H$\beta$ line to the blue and the stronger telluric lines to the red. The cross-correlation functions (CCFs) showed that KIC\,11494130 is a sharp and single-lined star. Radial velocities were derived from Gaussian fits to the CCFs. Internal fit errors were of the order of 25 m/s. 

We used the GSSP program \citep{2015A&A...581A.129T} that is based on the spectrum synthesis method to analyze the spectra, using $\lambda= 4720-5670$\,\AA. Synthetic spectra were computed with the aforementioned programs. We obtained $T_{\rm eff} = 6500\pm 100$\,K, $\log{g} = 3.85\pm 0.17$, [Fe/H] = $-0.33\pm0.06$, and $v\sin{i}=7.1\pm0.6$ km/s. The value of $T_{\rm eff}$ is lower than the 6750\,K given by \citet{2012ApJ...753...86T} and distinctly lower than the 7456\,K derived by \citet{2015ApJ...808..166S}. This difference in $T_{\rm eff}$ can be attributed to the use of different spectrum analysis methods and differences in spectral resolution. For the first time, we analyzed high-resolution spectra of the star and our model included lines from a broader spectral range. In particular the
simultaneous adjustment of $T_{\rm eff}$, $\log{g}$ and [Fe/H] has an influence on the determined temperature.

\section{Binary Models} \label{sec:model}

To model the light curve of the heartbeat star systems, we used the \texttt{ellc} package \citep{Maxted+16}. \texttt{ellc} includes an equilibrium polytrope treatment for stellar shape following \citet{Chandrasekhar+33}, which independently determines rotation and tidal distortion. For eccentric orbits, only equilibrium tides are accounted for and volume is assumed to be constant. Our models include the effects of gravity darkening, limb darkening, reflection, and Doppler boosting. For gravity darkening, \texttt{ellc} assumes that specific intensity is related to the local gravity by a power law with a gravity darkening exponent coefficient \citep{VonZeiepel24}. To correct for the effect of viewing angle on specific intensity, we adopted the linear limb darkening law from \cite{Maxted+16} and \cite{Schwarzschild+06}. Additionally, reflection accounts for irradiation from the companion star, and \texttt{ellc} adopts a simple irradiation model with three parameters that are related to the angular dependence and stellar surface distribution of specific intensity \citep{Maxted+16}. We adopted Doppler boosting factors following \citet{Placek+19}, and appropriate limb darkening and gravity darkening coefficients from \citet{Claret+11}.

Final models for our systems were obtained using a combination of \texttt{ellc} and \texttt{emcee}, a Markov chain Monte Carlo (MCMC) algorithm \citep{Foreman-Mackey+13}.
Our MCMC modeling fitted for 10 parameters: the primary radius $r_1$, secondary radius $r_2$, semi major axis $a$, mass ratio (secondary mass divided by primary mass) $q$, inclination $i$, eccentricity $e$, argument of periastron $\omega$, surface brightness ratio (secondary divided by primary) $J$, time of mid-eclipse $t_0$ (or luminosity minimum when no eclipse is present), and radial velocity offset $v_0$.

The likelihood of the light curve fit was determined following \citet{Burdge+19} by first calculating the difference between the light curve model and the \textit{Kepler} light curve data, and then finding the normalized logarithm of the probability density function of this difference. The likelihood of the radial velocity fits were similarly determined, using the difference between the radial velocity model and data for each star. The total model fit likelihood was the sum of the light curve likelihood and the radial velocity likelihoods.

\subsection{KIC 6117415 model} \label{sec:6117model}
We constrained the parameters $a$, $q$, $e$, and $\omega$ through radial velocity fitting described in Section~\ref{sec:model}. For our binary model fit that simultaneously fits the light curve and radial velocity data, we adopt Gaussian likelihoods for $a$, $q$, $e$, and $\omega$ centered about their estimated values previously determined through radial velocity fitting. For all other parameters, we selected flat, uniform priors with lower and upper boundaries set so as to avoid unphysical models. We adopted a fixed period of $P = 19.7416$~days, consistent with the \textit{Kepler} catalog.

For gravity darkening, we fixed the exponent to be $0.45$, consistent with \citet{Claret+11} for the surface temperature of the stars, which have thin convective envelopes. The limb darkening coefficient was set to be $0.7$ from \citet{Claret+11}. For reflection, we fixed the geometric albedo for both stars to $0.6$, the theoretical value for stars with convective envelopes \citep{Rucinski+69, Beck+14}. Since KIC 6117415 has constituent stars of similar mass, temperature, and radial velocity, the effect of Doppler boosting is negligible and thus we elected not to fit for Doppler boosting effects.

\begin{figure}
    \centering{\includegraphics[width=0.87\columnwidth]{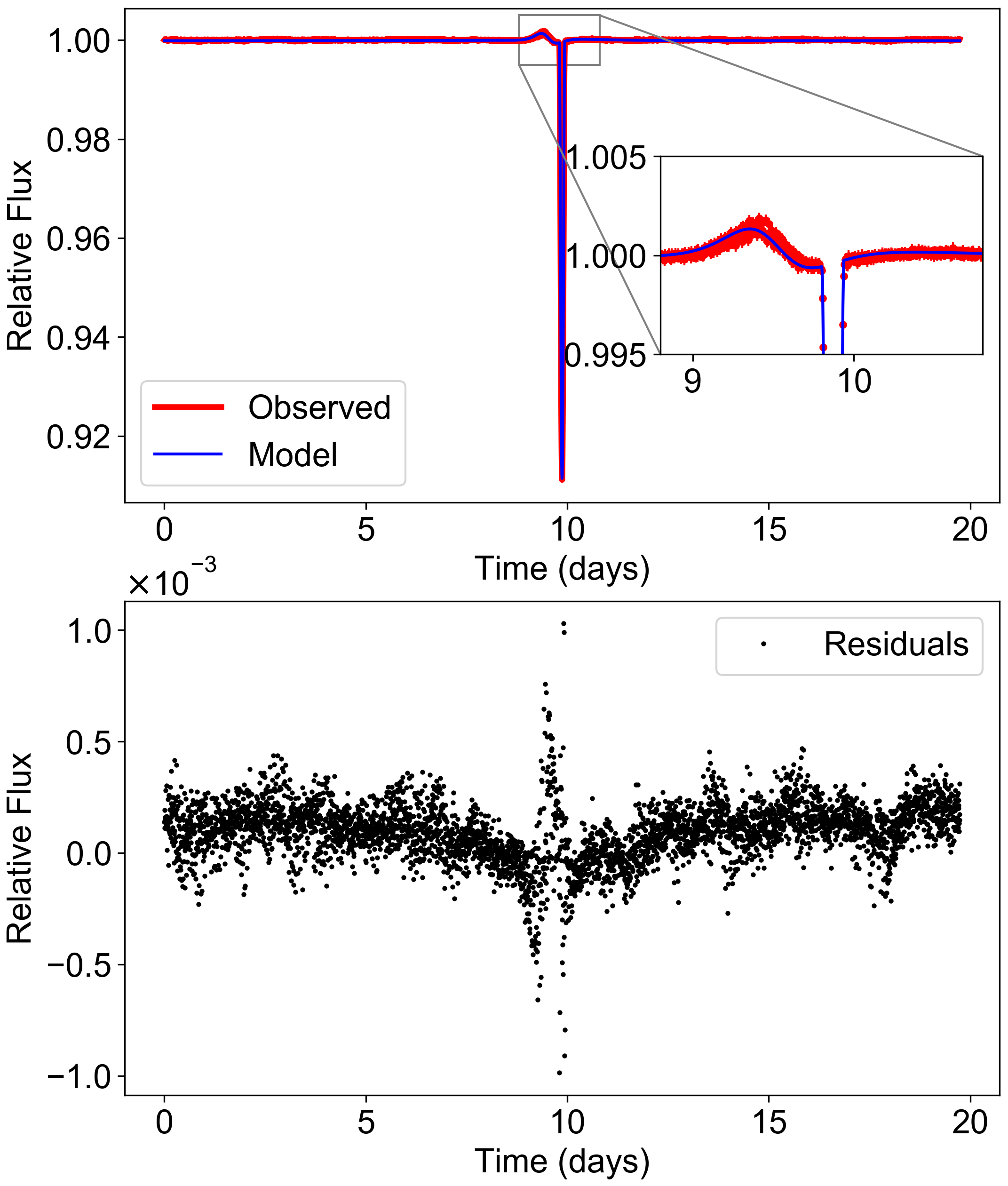}}
	\caption{\textbf{Light curve model of KIC 6117415.} Upper panel: the best fit light curve (blue) to the detrended, folded, and binned \textit{Kepler} data of KIC 6117415 (red) obtained through the process described in Section~\ref{sec:data}. The inset highlights the ellipsoidal heartbeat modulation near periastron. Bottom panel: residuals of the best fit model.}
	\label{fig:6117lc}
\end{figure}

\begin{figure}
	\centering{\includegraphics[width=0.87\columnwidth]{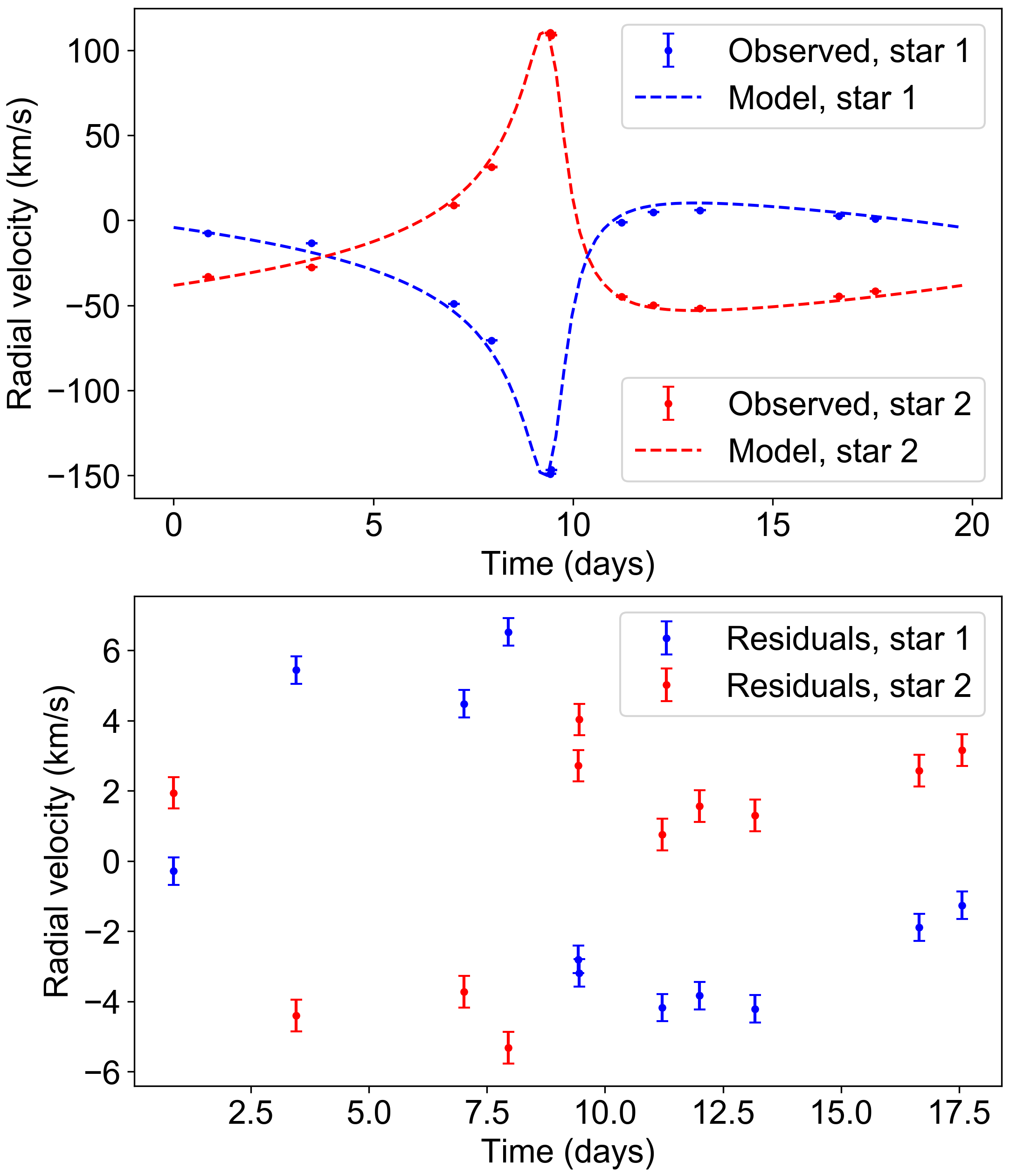}}
	\caption{\textbf{Radial velocities of KIC 6117415.} Upper panel: the best fit radial velocity curves (dashed lines) to the radial velocity data of KIC 6117415 (points) obtained through binary modeling. The primary star is in blue, and the secondary star is in red. Bottom panel: residuals of the best fit models.}
	\label{fig:6117rv}
\end{figure}

The best fit light curve and radial velocity curves are presented in Figure~\ref{fig:6117lc} and Figure~\ref{fig:6117rv}. Both the light curve and radial velocity curves are fairly well matched, with residuals much smaller than the measured flux and radial velocity variations. However, we note that the residuals are larger than the small measurement uncertainties, so our models are not fully consistent with the data. The increase in residuals during the eclipse arises from imperfect eclipse modeling. This is partly due to an imperfect limb-darkening model, since different limb-darkening coefficients and models produce differences in the amount of light blocked during the eclipse, which can lead to eclipse residuals. Additionally, the precise photometric measurements of Kepler have revealed the overall limitations in binary light curve synthesis codes such as \texttt{ellc} in its treatment of other parameters such as albedo. Our fitted masses are $m_1 = 1.461 \pm 0.006 ~\text{M}_\odot$ and $m_2 = 1.426 \pm 0.006 ~\text{M}_\odot$, with radii $r_1 = 1.462 \pm 0.004~\text{R}_\odot$ and $r_2 = 1.100 \pm 0.001 ~\text{R}_\odot$. Posteriors for the fitted parameters are presented in Section~\ref{appendix} (Figure~\ref{fig:6117corner}), and all parameters converged to approximately Gaussian distributions. The fitted values of all fitted and derived parameters are presented in Table~\ref{table:6117mcmc}.

We note that the uncertainties presented above are from MCMC fitting and are very underestimated compared to the true uncertainty of physical quantities. The residuals are not Gaussian due to the imperfect physics included in our binary models, and it is these systematics that are responsible for our measurement errors and uncertainties. This applies to all three systems discussed in this work. To better quantify the true uncertainties, we refit our system with geometric albedos of $0.5$ and $0.7$ and found that the best fit masses and radii differed by $0.06~\text{R}_\odot$, $0.05~\text{R}_\odot$, $0.05~\text{M}_\odot$, and $0.04~\text{M}_\odot$ for $r_1$, $r_2$, $m_1$, and $m_2$ respectively. Since geometric albedos of $0.5$ and $0.7$ are representative of the range of realistic values for the albedo for convective stars, these estimated uncertainties more fully capture the realistic uncertainties of our fitted values. The uncertainties in Table~\ref{table:6117mcmc} accounts for both the uncertainties from MCMC fitting and the estimated uncertainties from refitting with different geometric albedos. These larger uncertainties also help alleviate the concern of apparently different stellar radii despite the nominally similar stellar masses.

\begin{table}
\hspace{1.5em}\begin{tabular}{ll}
\hline
\noindent Parameter & Fitted value \\
\hline
Primary radius, $r_1$ (R$_\odot$)                      & $1.46 \pm 0.06$       \\
Secondary radius, $r_2$ (R$_\odot$)                    & $1.10 \pm 0.05$       \\
Semi major axis, $a$ (R$_\odot$)                        & $43.79 \pm 0.02$        \\
Primary mass, $m_1$ (M$_\odot$)               & $1.46 \pm 0.05$       \\
Secondary mass, $m_2$   (M$_\odot$)          & $1.43 \pm 0.04$       \\
Inclination, $i$ (degrees)                             & $83.16 \pm 0.01$       \\
Argument of periastron, $\omega$ (radians)             & $3.720 \pm 0.001$       \\
Eccentricity, $e$                                      & $0.7343 \pm 0.0004$      \\
Surface brightness ratio, $J$ & $0.800 \pm 0.003$     \\
Time of mid-eclipse, $t_0$ (days)                      & $7.795 \pm 0.003$       \\
Radial velocity offset, $v_0$ (km/s)                   & $-20.98 \pm 0.02$ \\
\hline\\
\end{tabular}

\begin{centering}
\begin{tabular}{lll}
\hline
Parameter & Primary star   & Secondary star\\
\hline
$T_{\rm eff}$ (K) & $6110 \pm 64$  & $6026 \pm 60$ \\
$\log g$ (cgs) & $3.97 \pm 0.05$ & $4.09 \pm 0.06$\\
$v\sin i$ (km s$^{-1}$) & $19.0 \pm 0.3$ & $20.2 \pm 0.5$\\
$[\textrm{Fe/H}]$ (dex) & $-0.60 \pm 0.05$ & $-0.74 \pm 0.05$
\footnote{We obtained similar results (within one sigma) if imposing the two stars have the same $[\textrm{Fe/H}]$}\\
\hline
\end{tabular}
\end{centering}

\begin{centering}
\hspace{0.5em}\begin{tabular}{lc}
\\
\hline
Parameter & SpecMatch best fit value \\
\hline
$T_{\rm eff}$ (K) & $6392 \pm 100$  \\
$\log g$ (cgs) & $4.41 \pm 0.10$\\
$v\sin i$ (km s$^{-1}$) & $19.65 \pm 1.0$ \\
$[\textrm{Fe/H}]$ (dex) & $-0.10 \pm 0.06$ \\
\hline
\end{tabular}
\end{centering}
\caption{Upper: The best fit parameters for the light curve and radial velocity data for KIC 6117415. Uncertainties were determined using a combination of $1\sigma$ fitting uncertainties and estimated uncertainties from refitting with different geometric albedos (see text).
Middle: The best fit parameters for the Keck/HIRES spectroscopy data. Lower: Parameters obtained from SpecMatch (Synthetic).}
\label{table:6117mcmc}
\end{table}

We compared our best fit parameters with luminosity and temperature constraints derived from \texttt{Gaia} data release 2 presented in \citet{Berger+18}. The estimated temperature of KIC 6117415 in the SpecMatch HIRES California Planet Search database is $T_{\text{eff}} = 6392 \pm 100$~K, and is roughly consistent with our mass estimate for a main sequence star such as KIC 6117415. Using our best fit radii and temperature from SpecMatch HIRES, we estimate (via the Stefan-Boltzmann equation) a total system luminosity of $4.9 \pm 0.5 ~ L_{\odot}$. This is consistent with the luminosity of $5.4 \pm 0.1 ~ L_{\odot}$ reported in \citet{Berger+18}.

\begin{figure}
	\includegraphics[width=\columnwidth]{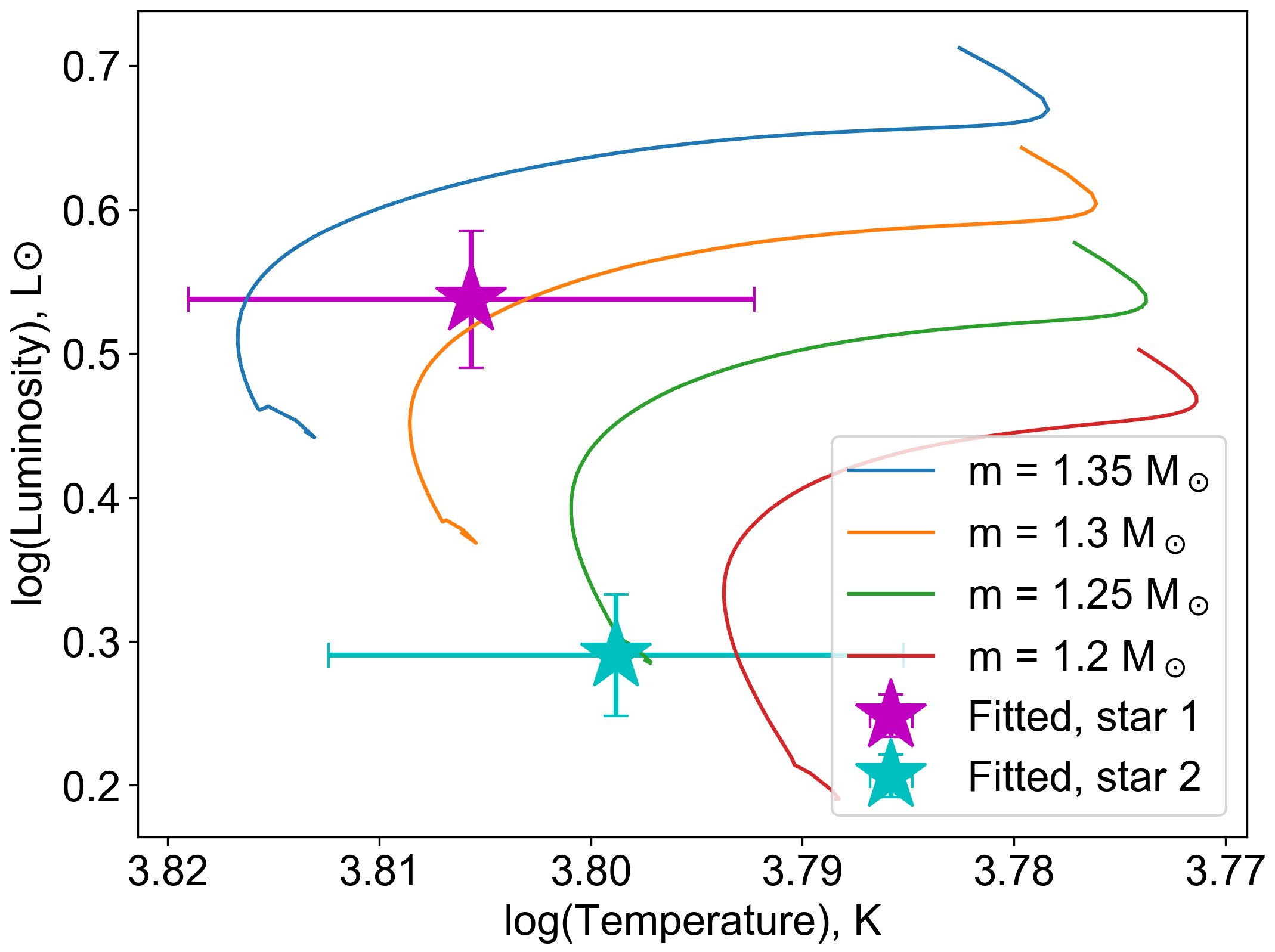}
	\caption{\textbf{Hertzsprung-Russell diagram of \texttt{MESA} models for KIC 6117415.} Each line is a \texttt{MESA} evolutionary track for a star with metallicity of $[\textrm{Fe/H}]= -0.1$. The magenta star marks the SpecMatch HIRES temperature of $6392 \pm 200$~K and inferred luminosity for the primary star, and the cyan star marks the inferred properties of the secondary star (see text).}
	\label{fig:6117HR}
\end{figure}

We created stellar evolution models for KIC 6117415 using \texttt{MESA} \citep{Paxton+11,paxton:13,paxton:15,paxton:18,paxton:19} with the SpecMatch metallicity and a range of masses within $10\%$ of our best fit masses. Such models are required in order to calculate predictions for TEOs (see Section~\ref{sec:TEO}). Model evolution tracks are compared with temperature and luminosity measurements in Figure \ref{fig:6117HR}. We note that due to the conflicting temperatures for KIC 6117415 between those from SpecMatch and those derived from fits to synthetic spectra (see Section~\ref{sec:keck}), the luminosities shown in Figure~\ref{fig:6117HR} are uncertain. To reflect this mismatch, we increase the temperature uncertainty from $100~$K to $200~$K. In Figure \ref{fig:6117HR}, the individual luminosities for the stars were estimated using the total system luminosity of $5.4 \pm 0.1 ~ L_{\odot}$, a $T_{\rm eff}$ difference of 100 K, and using the ratio between the radius of the primary and secondary stars obtained from light curve fitting. We note that the best-fit evolutionary models imply masses of $\approx 1.3~\text{M}_\odot$, whereas the masses obtained from light curve fitting are about 10\% larger. This slight tension likely arises from imperfect light curve and radial velocity modeling, as evidenced from the offset in Figure \ref{fig:6117rv} that suggests a mass function overestimated by several percent. 

\subsection{KIC 11494130 model}
For our binary model fit of KIC 11494130, we used radial velocity data from \citet{Smullen+15} and the TLS observatory (see Section~\ref{sec:TLS}). We adopted flat, uniform priors with lower and upper boundaries set so as to avoid unphysical models for all parameters. We used the \textit{Kepler} period of $P = 18.9554$ days, and gravity and limb darkening coefficients of $0.6$ and $0.8$. While we adopted a geometric albedo of $0.6$ for the primary star following \citet{Rucinski+69} and \citet{Beck+14}, our secondary star is a M-dwarf and may have a lower geometric albedo since its flux may be radiated out in the infrared rather than the visible. We selected a geometric albedo of $0.2$ for the secondary star, and note that the shape of the light curve near periastron is ultimately insensitive to different geometric albedo values. We selected a Doppler boosting factor of $3.7$ following \citet{Placek+19}.

\begin{figure}
	\centering{\includegraphics[width=0.87\columnwidth]{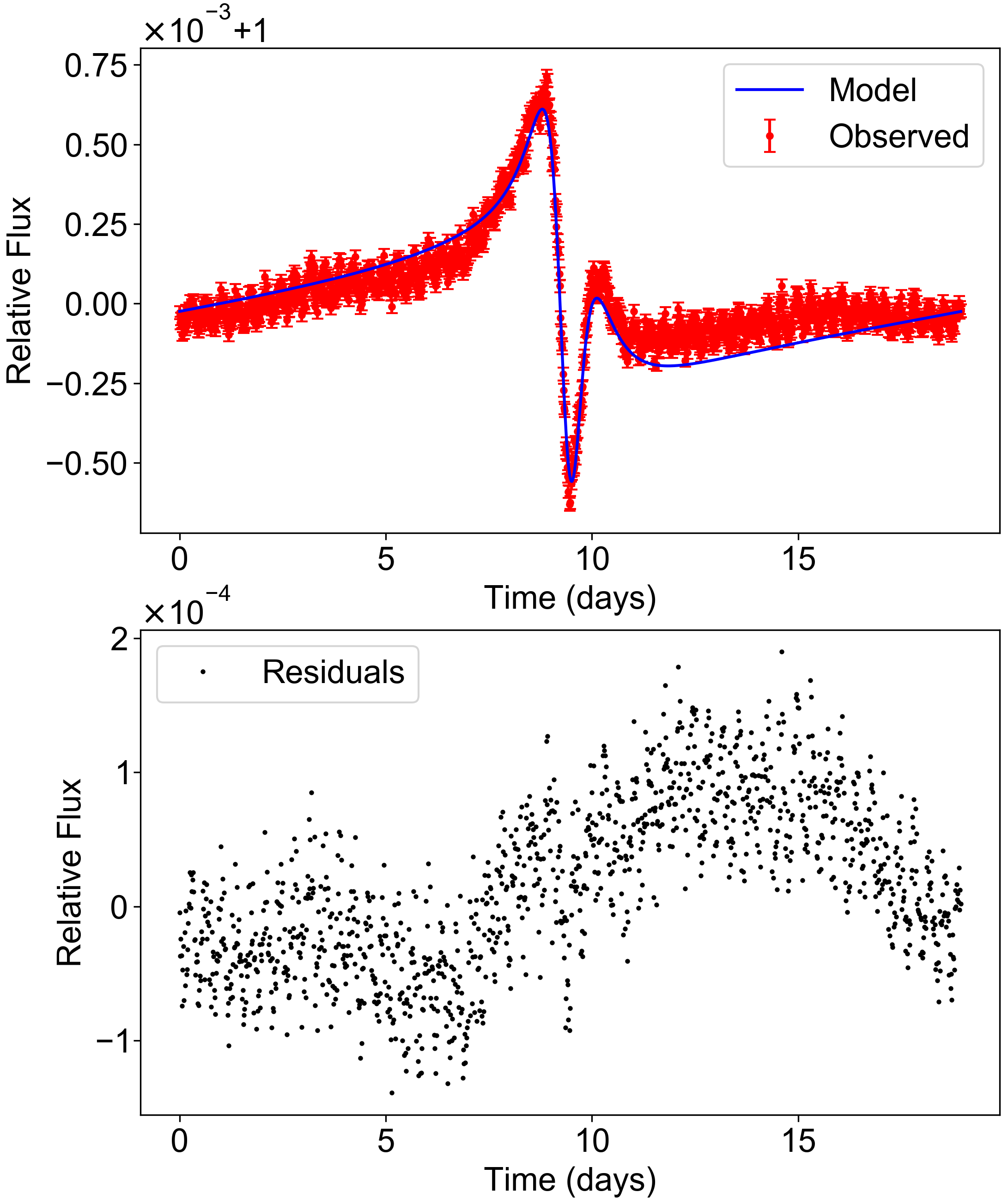}}
	\caption{\textbf{Light curve model of KIC 11494130.} Upper panel: the best fit light curve (blue) to the detrended, folded, and binned \textit{Kepler} data of KIC 11494130 (red) obtained through the process described in Section~\ref{sec:data}. Bottom panel: residuals of the best fit model.}
	\label{fig:114lc}
\end{figure}

\begin{figure}
	\centering{\includegraphics[width=0.87\columnwidth]{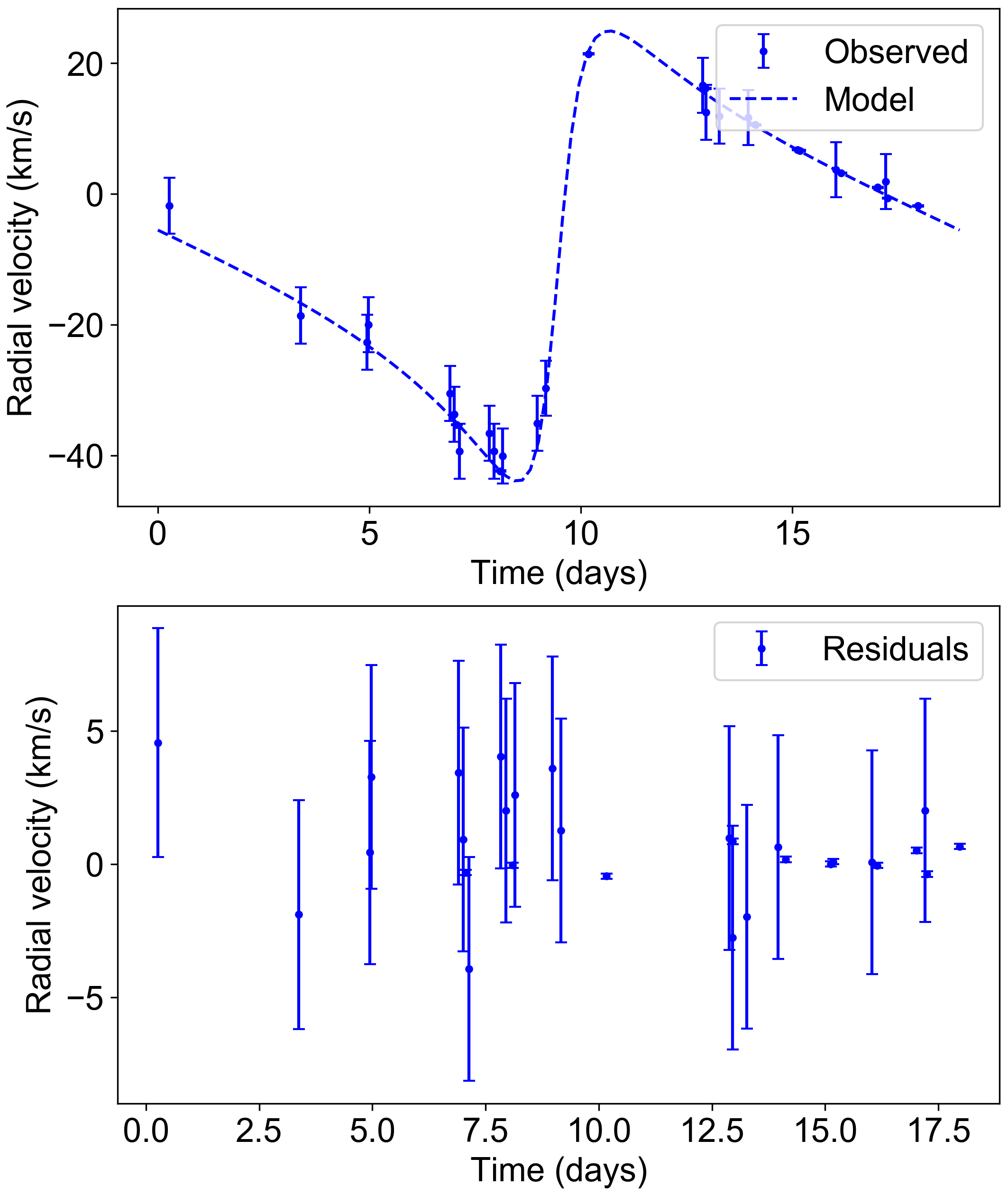}}
	\caption{\textbf{Radial velocities of KIC 11494130.} Upper panel: the best fit radial velocity curve (dashed line) to the radial velocity data of KIC 11494130 (points) obtained through binary modeling. Bottom panel: residuals of the best fit model.}
	\label{fig:114rv}
\end{figure}

Figures~\ref{fig:114lc} and \ref{fig:114rv} show the best fit light curve and radial velocity curves of KIC 11494130. As with KIC 6117415, the posteriors converged to approximately Gaussian distributions (see Figure~\ref{fig:114corner} in Section~\ref{appendix}). The trend in the light curve fit residuals is likely due to imperfect detrending of the light curve data. Specifically, the light curve data away from periastron depends heavily on the choice of polynomial order, tolerance, and window length when applying the Savitzky-Golay filter during detrending.

\begin{table}
\hspace{0.5em} \begin{tabular}{ll}
\hline
Parameter                                  & Fitted value \\
\hline
Primary radius, $r_1$ (R$_\odot$)                      & $1.59 \pm 0.06$       \\
Secondary radius, $r_2$ (R$_\odot$)                    & $0.65 \pm 0.05$       \\
Semi major axis, $a$ (R$_\odot$)                        & $36.79 \pm 0.04$        \\
Primary mass, $m_1$ (M$_\odot$)               & $1.35 \pm 0.02$       \\
Secondary mass, $m_2$   (M$_\odot$)          & $0.50 \pm 0.02$       \\
Inclination, $i$ (degrees)                             & $79.20 \pm 0.09$       \\
Argument of periastron, $\omega$ (radians)             & $4.59 \pm 0.01$       \\
Eccentricity, $e$                                      & $0.66 \pm 0.03$      \\
Surface brightness ratio, $J$ & $0.3 \pm 0.1$     \\
Time of mid-eclipse, $t_0$ (days)                      & $1.35 \pm 0.07$       \\
Radial velocity offset, $v_0$ (km/s)                   & $-6.79 \pm 0.03$ \\
\hline\\
\end{tabular}

\noindent\begin{tabular}{lc}
\hline
Parameter & TLS best fit value \\
\hline
$T_{\rm eff}$ (K) & $6500 \pm 110$  \\
$\log g$ (cgs) & $3.85 \pm 0.17 $\\
$v\sin i$ (km s$^{-1}$) & $7.1 \pm 0.6$ \\
$[\textrm{Fe/H}]$ (dex) & $-0.33 \pm 0.06$ \\
\hline\\
\end{tabular}

\noindent\begin{tabular}{lc}
\hline
Parameter & SpecMatch best fit value \\
\hline
$T_{\rm eff}$ (K) & $6599 \pm 110$  \\
$\log g$ (cgs) & $4.23 \pm 0.10 $\\
$v\sin i$ (km s$^{-1}$) & $0.1 \pm 1.0$ \\
$[\textrm{Fe/H}]$ (dex) & $-0.05 \pm 0.09$ \\
\hline
\end{tabular}

\caption{The best fit parameters for the light curve, radial velocity, and spectral data for KIC 11494130. The format follows Table~\ref{table:6117mcmc}.}
\label{table:114mcmc}
\end{table}

Our fitted masses are $m_1 = 1.35 \pm 0.02 ~\text{M}_\odot$ and $m_2 = 0.50 \pm 0.02 ~\text{M}_\odot$, with radii $r_1 = 1.59 \pm 0.06 ~\text{R}_\odot$ and $r_2 = 0.65 \pm 0.05 ~\text{R}_\odot$. Similar to the case for KIC 6117415, we refit our system using geometric albedos of $0.5$ and $0.7$ for the primary star and found small differences of $0.01~R_\odot$ for $r_1$ and $r_2$ and $0.01~M_\odot$ for $m_1$ and $m_2$. Table~\ref{table:114mcmc} presents our final values of all fitted and derived parameters, with uncertainties that account for both fitting uncertainties and estimated uncertainties from varying the geometric albedo of the primary star. With an effective temperature of $6500\pm110~$K from TLS spectra, we compute a system luminosity of $4.7 \pm 0.3 ~ L_\odot$, consistent with the luminosity of $4.5 \pm 0.3 ~ L_\odot$ from \citet{Berger+18}. We note that, here, the effective temperature derived from TLS spectra is consistent with the temperature from SpecMatch. We use the temperature derived from TLS spectra for our analysis. 

We also note that our eccentricity measurement of $e= 0.66 \pm 0.03$ is significantly different from that presented in \citet{Smullen+15}, who found $e = 0.49 \pm 0.05$.
This difference may be due to the additional radial velocity data we used from the TLS observatory that better constrains the shape of the radial velocity curve. Specifically, the radial velocity data point at $\sim 10~$days was obtained from the TLS observatory and unavailable to \citet{Smullen+15}. Without a constraint near periastron, their radial velocity fit resulted in a curve different in shape and orbital solution from the one presented here.

\begin{figure}
	\includegraphics[width=\columnwidth]{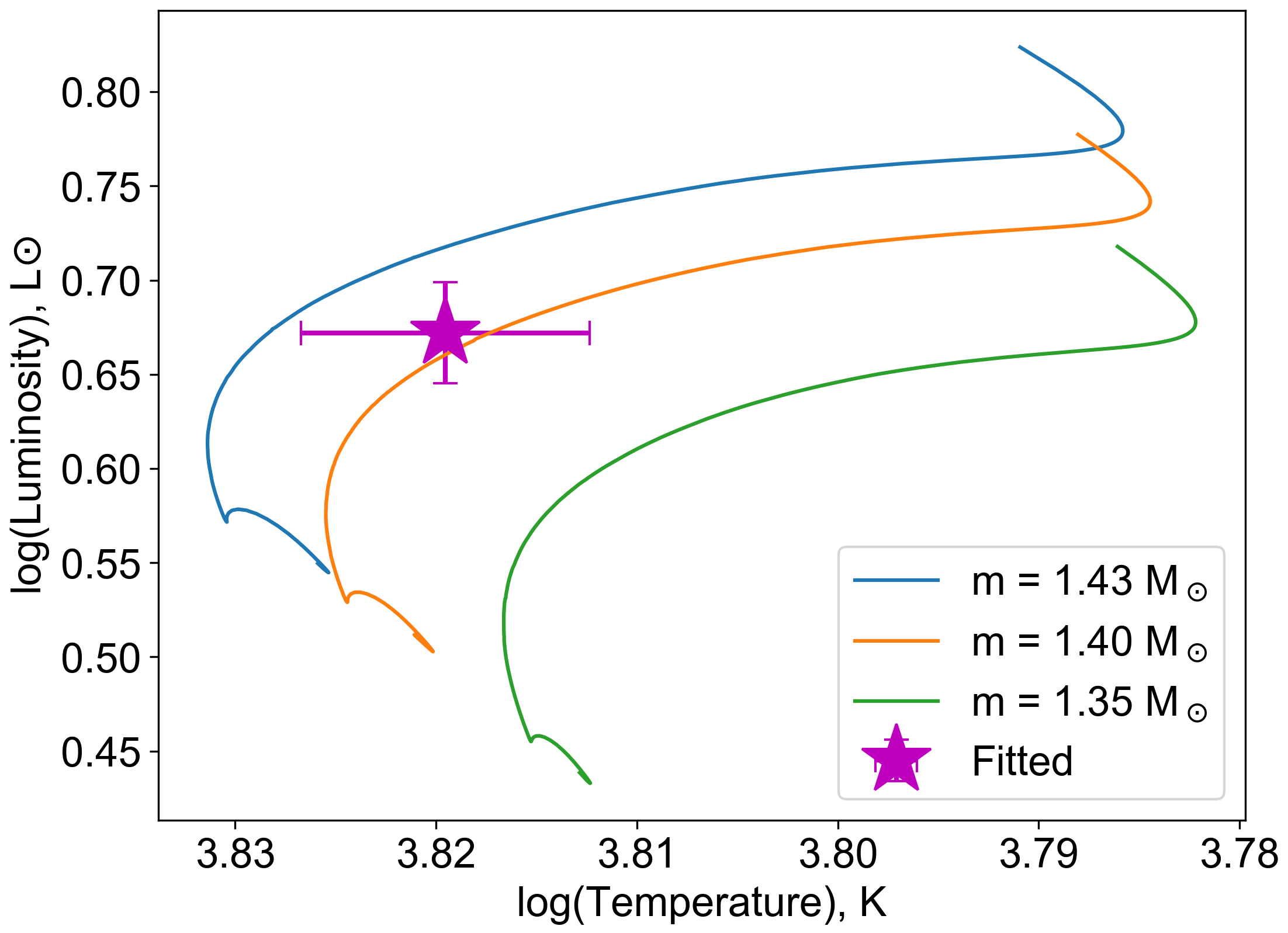}
	\caption{\textbf{Hertzsprung-Russell diagram of \texttt{MESA} models for KIC 11494130.} We modeled each evolutionary track using a metallicity of $[\textrm{Fe/H}] = -0.33$ following the TLS spectral analysis. The magenta star marks the TLS spectra temperature of $6500 \pm 110$~K and subsequent luminosity of $4.7 \pm 0.3 ~ L_\odot$, consistent with \citet{Berger+18}.}
	\label{fig:114HR}
\end{figure}

Figure \ref{fig:114HR} compares MESA model tracks to our measured temperature and luminosity for the primary. The \texttt{MESA} models use the somewhat low metallicity from TLS spectra. The best fit models of $\approx 1.4~\text{M}_\odot$ are within a few percent of masses based on light curve modeling, indicating a satisfactory agreement.

\subsection{KIC 5790807 model}

We created a binary model using radial velocity data from \citet{Shporer+16}. We used flat uniform priors for all parameters with bounds set to ensure physical models. We used the \textit{Kepler} period of $79.996$ days, and coefficients for gravity darkening, limb darkening, and Doppler boosting were $0.6$, $0.8$, and $3.5$. Similar to KIC 11494130, the secondary star is an M-dwarf with small geometric albedo. Therefore, we adopted a geometric albedo of $0.6$ for the primary star following \citet{Rucinski+69} and \citet{Beck+14} and $0.2$ for the secondary. We note that the shape of the light curve near periastron is weakly dependent on the secondary star's geometric albedo.

\begin{figure}
	\centering{\includegraphics[width=0.87\columnwidth]{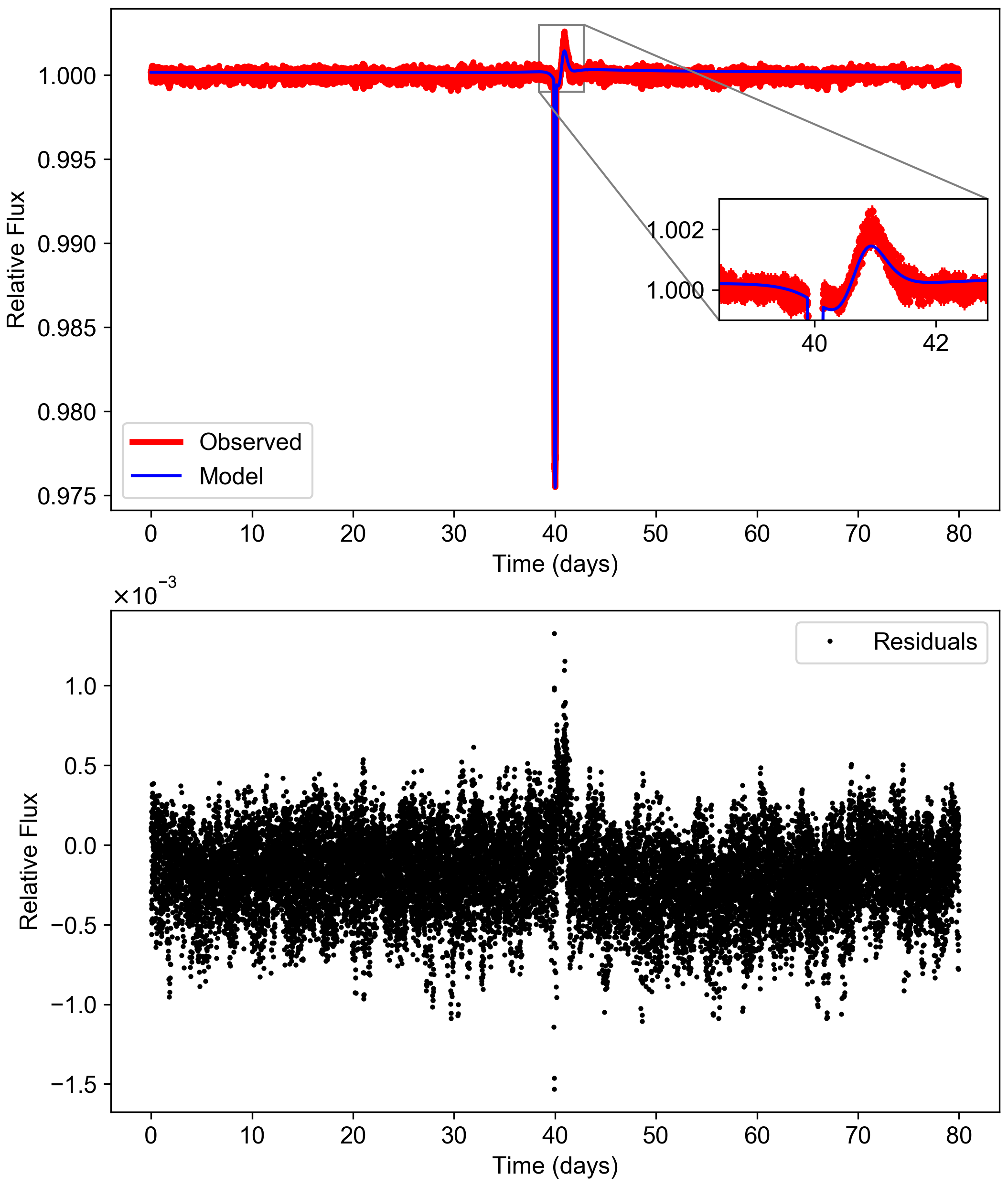}}
	\caption{\textbf{Light curve model of KIC 5790807.} Upper panel: the best fit light curve (blue) to the detrended, folded, and binned \textit{Kepler} data of KIC 5790807 (red) obtained through the process described in Section~\ref{sec:data}. Bottom panel: residuals of the best fit model.}
	\label{fig:57lc}
\end{figure}

\begin{figure}
	\centering{\includegraphics[width=0.87\columnwidth]{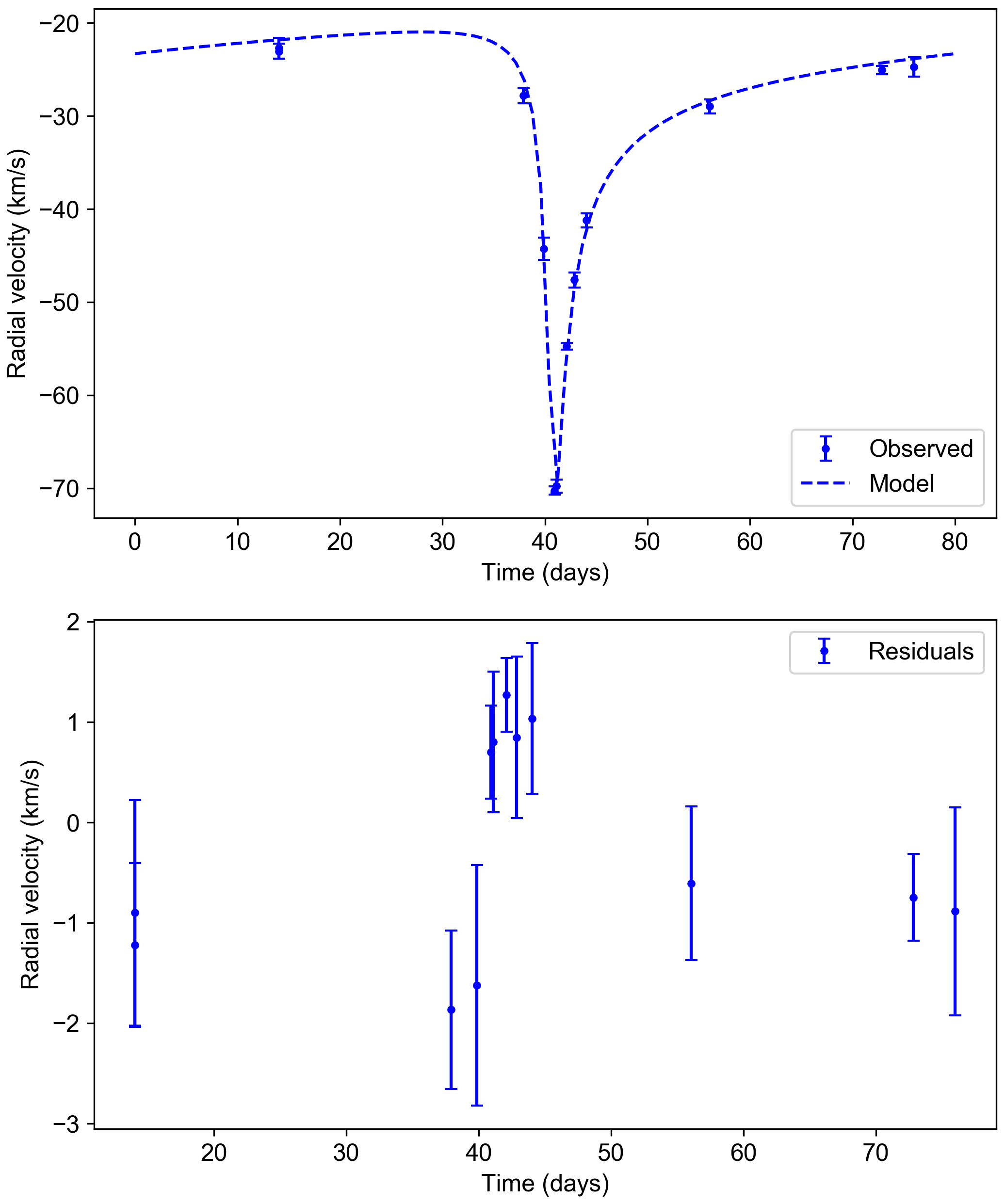}}
	\caption{\textbf{Radial velocities of KIC 5790807.} Upper panel: the best fit radial velocity curve (dashed line) to the radial velocity data of KIC 5790807 (points) obtained through binary modeling. Bottom panel: residuals of the best fit model.}
	\label{fig:57rv}
\end{figure}

The best fit light curve, radial velocity curve, fitted values of parameters are shown in Figure~\ref{fig:57lc}, Figure~\ref{fig:57rv}, and Table~\ref{table:57mcmc}. The uncertainties presented in Table~\ref{table:57mcmc} follows those from Table~\ref{table:114mcmc}, with estimated uncertainties derived from refitting our system with geometric albedos of $0.5$ and $0.7$ for the primary star. We found that the radii and masses differed by $0.01~\text{R}_\odot$, $0.002~\text{R}_\odot$, $0.02~\text{M}_\odot$, and $0.01~\text{M}_\odot$ for $r_1$, $r_2$, $m_1$, and $m_2$ respectively. Posteriors are shown in Figure~\ref{fig:57corner} and are approximately Gaussian. Our estimated total system luminosity of $12 \pm 1 ~L_\odot$ is lower than the luminosity of $14.6 \pm 0.4 ~ L_\odot$ from \citet{Berger+18}. This could be largely due to a difference of around $300$~K between the temperature quoted in \citet{Berger+18} ($6786 \pm 136$~K) as compared to the SpecMatch HIRES data we have adopted ($6466 \pm 110$~K) in our calculation of total system luminosity. However, since the lower limit of the temperature from \citet{Berger+18} $6650~$K is comparable to the upper limit of the SpecMatch HIRES temperature $6576~$K, we accept our fit results for KIC 5790807. The different temperatures may arise from differences between the photometric estimate from \citet{Berger+18}, and the spectroscopic technique from SpecMatch.

\begin{figure}
	\includegraphics[width=\columnwidth]{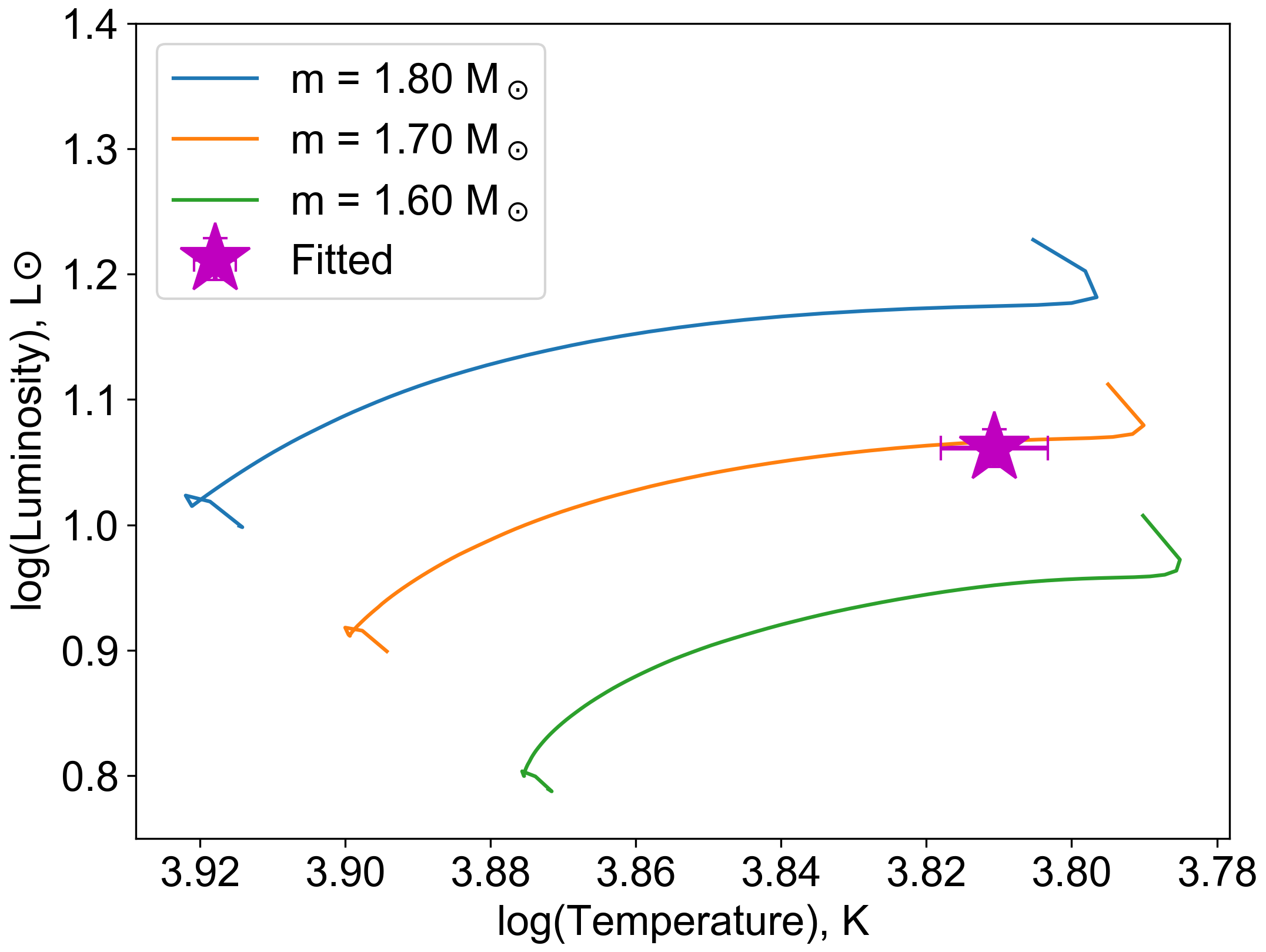}
	\caption{\textbf{Hertzsprung-Russell diagram of \texttt{MESA} models for KIC 5790807.}  We modeled each evolutionary track using a metallicity of $[\textrm{Fe/H}] = 0.06$ following the SpecMatch results. The magenta star marks the SpecMatch HIRES temperature of $6466$~K and the inferred luminosity of $12 \pm 1 ~ L_\odot$.}
	\label{fig:57HR}
\end{figure}

Figure \ref{fig:57HR} compares \texttt{MESA} models (with the SpecMatch metallicity) to our measured temperature and luminosity of KIC 5790807. We find a satisfactory match for models with masses of $\approx 1.7~\text{M}_\odot$, within a few percent of the mass inferred from light curve modeling. The relatively large radius and cool temperature of the primary indicates it is nearing the end of its main sequence evolution.  

\begin{table}

\hspace{1em}\begin{tabular}{ll}
\hline
Parameter                                  & Best fit value \\
\hline
Primary radius, $r_1$ (R$_\odot$)                      & $2.68 \pm 0.02$       \\
Secondary radius, $r_2$ (R$_\odot$)                    & $0.397 \pm 0.003$       \\
Semi major axis, $a$ (R$_\odot$)                        & $101.36 \pm 0.06$        \\
Primary mass, $m_1$ (M$_\odot$)               & $1.74 \pm 0.02$       \\
Secondary mass, $m_2$   (M$_\odot$)          & $0.44 \pm 0.01$       \\
Inclination, $i$ (degrees)                             & $85.82 \pm 0.01$       \\
Argument of periastron, $\omega$ (radians)             & $2.715 \pm 0.004$       \\
Eccentricity, $e$                                      & $0.855 \pm 0.001$      \\
Surface brightness ratio, $J$ & $0.48 \pm 0.03$     \\
Time of mid-eclipse, $t_0$ (days)                      & $40.009 \pm 0.009$       \\
Radial velocity offset, $v_0$ (km/s)                   & $-26.56 \pm 0.02$ \\
\hline
\\
\end{tabular}

\begin{tabular}{lc}
\hline
Parameter & SpecMatch best fit value \\
\hline
$T_{\rm eff}$ (K) & $6466 \pm 110$  \\
$\log g$ (cgs) & $3.42 \pm 0.10$\\
$v\sin i$ (km s$^{-1}$) & $14.56 \pm 1.0$ \\
$[\textrm{Fe/H}]$ (dex) & $0.06 \pm 0.09$ \\
\hline
\end{tabular}

\caption{The best fit parameters for the light curve and radial velocity data for KIC 5790807. The format follows Tables~\ref{table:6117mcmc} and \ref{table:114mcmc}.}
\label{table:57mcmc}
\end{table}

\newpage

\section{Frequency Analysis} \label{sec:freq}

To determine whether our systems exhibit any TEOs, we perform a frequency analysis on the phased light curve fit residuals shown in Figures~\ref{fig:6117lc}, \ref{fig:114lc}, and \ref{fig:57lc} in which we subtracted the binary model from the corresponding \textit{Kepler} light curve data. We conducted the frequency analysis using a Lomb-Scargle periodogram. To ensure that the systems' imperfectly modeled eclipse in KIC 6117415 and KIC 5790807 do not impact the analysis, all data near eclipses were removed. Since the orbital period of the systems are much longer than the short duration of the eclipses, this removal does not prohibitively contribute to the window pattern in the frequency spectrum.

\subsection{KIC 6117415 analysis}
\begin{figure}
	\includegraphics[width=\columnwidth]{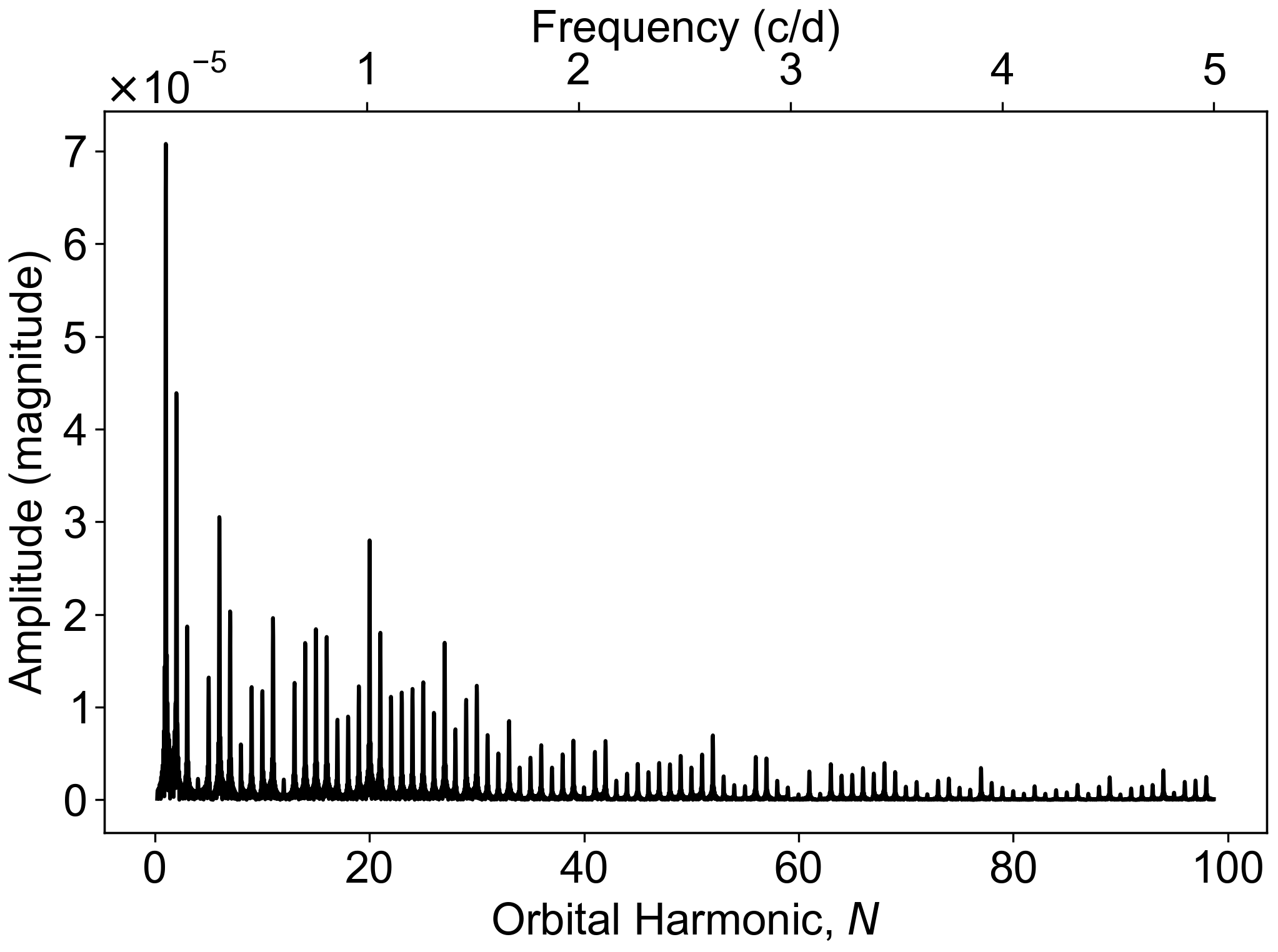}
	\caption{\textbf{Frequency spectrum of KIC 6117415 binary model residuals.} No high amplitude tidally excited oscillations (TEOs) are evident. All visible peaks occur at exact integer multiples (harmonics) of the orbital frequency of $5.07 \times 10^{-2}$~c/d, and are produced by TEOs, or are remnants of our imperfect light curve model. The peak at $N=20$ is unlikely to be a TEO (see text). }
	\label{fig:6117ft}
\end{figure}

We omitted data between $9.75$~days and $9.89$~days in our analysis to remove the effect of the eclipse residuals from our frequency analysis. We did not identify any high-amplitude TEOs in the frequency spectrum (Figure~\ref{fig:6117ft}). This was expected since KIC 6117415's \textit{Kepler} light curve does not visibly exhibit any large periodic oscillations throughout the orbit. While there is a noticeable peak at $N=20$ ($\simeq \! 1 \, {\rm d}^{-1}$), \cite{guo16} also found a peak in the unphased light curve of KIC 6117415 at 1.0 d$^{-1}$ at what appears to be the third harmonic of one of the stars' rotation frequency. Additionally, the amplitude of the peak at the $20^{\text{th}}$ orbital harmonic is sensitive to the choice of data omission near periastron. Thus, this peak is likely related to rotation or is an artefact from our imperfect light curve model rather than a signature of a TEO. We therefore conclude that KIC 6117415 does not exhibit any clear TEOs.

Using the frequency spectrum of KIC 6117415 as a guide, we consider possible TEOs as peaks with amplitude greater than $1 \times 10^{-5}$ and which are more than $\sim 2$ times higher in amplitude than neighboring peaks. This criteria, while somewhat arbitrary, appropriately omits some peaks that are due to imperfect light curve modeling such as the $N = 20$ peak in KIC 6117415.

We note that the observed peaks at very low and high orbital harmonics in all $3$ systems are artefacts of an imperfect light curve fit and can be considered as noise. At low $N$, the frequency spectra are especially noisy since our light curve fit does not perfectly capture the ellipsoidal modulation of light curves, which are typically peaked at low $N$. This is expected in all $3$ systems since low-frequency periodic variations can be seen in the light curve residuals in Figures~\ref{fig:6117lc}, \ref{fig:114lc}, and \ref{fig:57lc}.


\begin{figure}
	\includegraphics[width=\columnwidth]{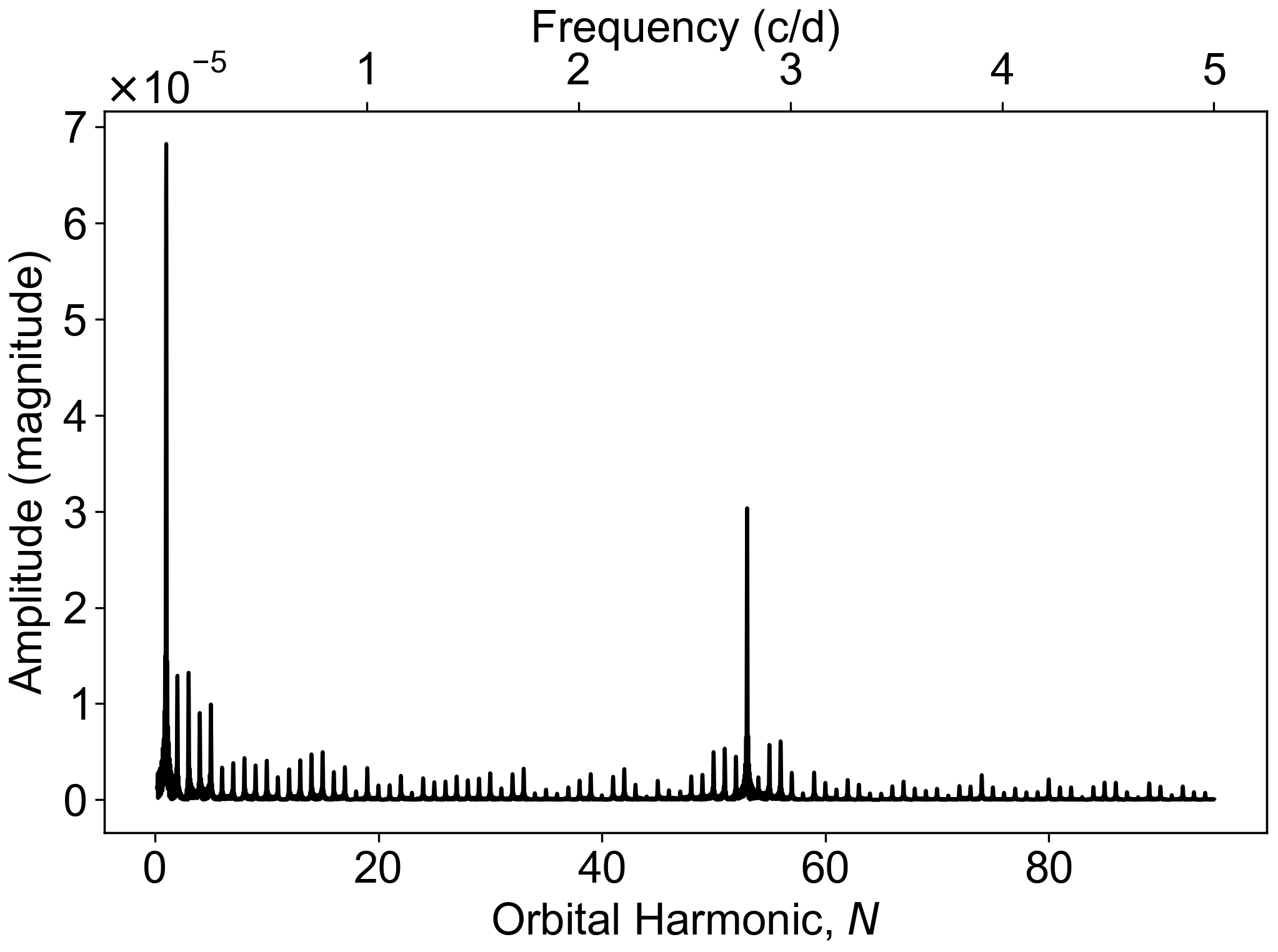}
	\caption{\textbf{Frequency spectrum of KIC 11494130 binary model residuals.} A prominent tidally excited oscillation occurs at $N=53$. The peak at $N=1$ is an artefact from imperfect light curve modeling. The peaks neighboring the prominent TEO at $N=53$ are likely due to the combination of frequencies from imperfect light curve modeling and the legitimate $N=53$ peak.}
	\label{fig:114ft}
\end{figure}

\subsection{KIC 11494130 analysis}

Since no eclipse is featured in the light curve of KIC 11494130, we used all residual data in our frequency analysis. As shown in Figure~\ref{fig:114ft}, a prominent TEO occurs at $N = 53$ (Figure~\ref{fig:114ft}). This is expected, since the \textit{Kepler} light curve data features visible TEOs throughout the orbit. We note that the presence of a high amplitude peak at $N=1$ is due to the imperfect light curve fit rather than a signature of a TEO. Other than the high-amplitude TEO at $N=53$, we do not identify any other clear TEOs.

\subsection{KIC 5790807 analysis}
\begin{figure}
	\includegraphics[width=\columnwidth]{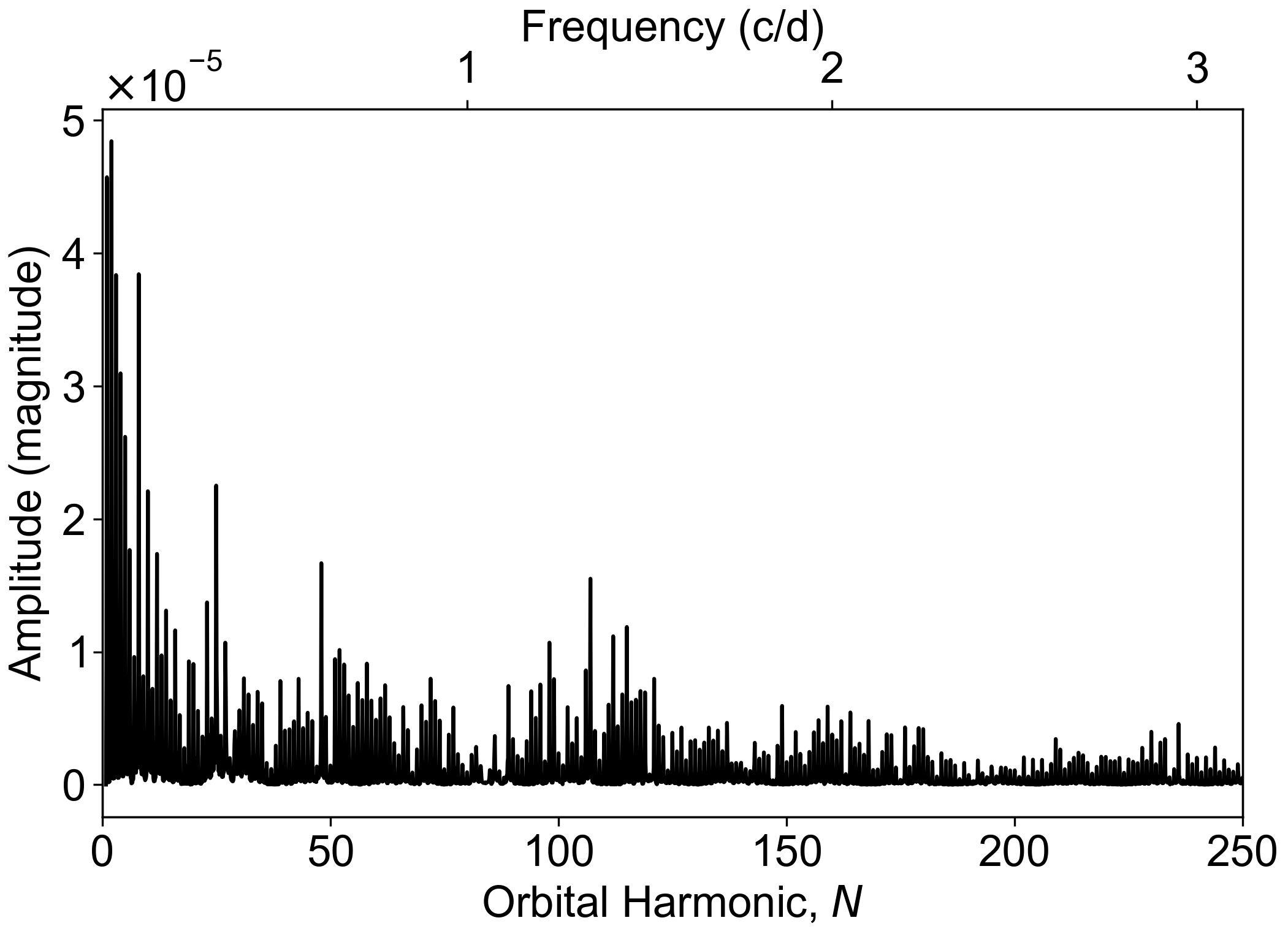}
	\caption{\textbf{Frequency spectrum of KIC 5790807 binary model residuals.} Likely tidally excited oscillations occur at $N= 48 \ \text{and } 107$, while the peaks below $N=30$ are likely artefacts from imperfect binary modeling.}
	\label{fig:57ft}
\end{figure}

Similar to KIC 6117415, we omitted data between $40.05$~days and $40.33$~days to remove the effect of the eclipse residuals from our analysis. As shown in Figure~\ref{fig:57ft}, we identify peaks at the $48^{\text{th}}$ and $107^{\text{th}}$ harmonics which may be TEOs. This is expected, since the presence of TEOs can be seen in the light curve. The cluster of peaks near $N=25$ and those below $N=10$ are likely artefacts from imperfect light curve modeling since their amplitudes are sensitive to the choice of data omission near periastron. Thus, these peaks are not considered probable TEOs.

\section{Tidally Excited Oscillations}
\label{sec:TEO}

\subsection{Tidal Models}\label{sec:tidalmodel}

TEOs are stellar oscillation modes, typically gravity modes (g modes), excited by the dynamic tidal forces throughout the binary star's orbit. \citet{Fuller+17} performed detailed calculations of the forced amplitudes, frequencies and phases of TEOs in eccentric binary systems, including a statistical approach with which to determine most probable TEO amplitudes. Here, we follow \citet{Fuller+17} and perform an analysis of the TEOs in KIC 6117415, KIC 11494130, and KIC 5790807.

The luminosity fluctuation due to a single oscillation mode (labeled $\alpha$) at orbital harmonic $N$ takes a sinusoidal form
\begin{equation}
    \frac{\Delta{L_N}}{L} \simeq A_N \sin(N\Omega t + \Delta_N) ,
\end{equation}
where $t$ is time from periastron, $\Omega = 2 \pi/P$ is the angular orbital frequency, $\Delta_N$ is the phase of surface luminosity perturbation relative to periastron, and the amplitude is

\begin{equation} \label{eq:A_N}
    A_N = \epsilon_l X_{Nm} V_{lm} |Q_\alpha L_\alpha| \frac{\omega_{Nm}}{\sqrt{(\omega_\alpha - \omega_{Nm})^2 + \gamma_\alpha^2}}.
\end{equation}

All terms appearing in Equation~\ref{eq:A_N} are defined in detail in \citet{Fuller+17}. The $\ell$ and $m$ subscripts are the multipole of the oscillation mode, with $l=2$ the most dominant in tidal excitation. $\omega_\alpha$ is the stellar oscillation mode frequency in the star's rotating frame of reference, and $\omega_{Nm}$ is the tidal forcing frequency in the rotating frame. $\epsilon_l$ is a dimensionless tidal forcing amplitude. For our best fit stellar/orbital parameters, we find $\epsilon_2 = 3.61 \times 10^{-5}$ for KIC 6117415, $\epsilon_2 = 2.85 \times 10^{-5}$ for KIC 11494130, and $\epsilon_2 = 4.69 \times 10^{-6}$ for KIC 5790807. 

The Hansen coefficient $X_{Nm}$ encapsulates the strength of the tidal forcing in an orbit with eccentricity $e$ at an orbital harmonic $N$. $V_{lm}$ corrects for mode visibility due to viewing angle and is found by $V_{lm} = |Y_{lm}(i_s,0)|$, where $Y_{lm}$ is a spherical harmonic and $i_s$ is the inclination between the star's rotation axis and the line of sight. $Q_\alpha$ is a dimensionless quadrupole moment that captures the spatial coupling between mode $\alpha$ and the tidal potential, $L_\alpha$ is the luminosity fluctuation produced by mode $\alpha$ at the stellar surface, and $\gamma_\alpha$ is the mode damping rate. The parameters $\omega_\alpha$, $Q_\alpha$, $L_\alpha$, and $\gamma_\alpha$ are dependent on stellar models and oscillation modes, while $\epsilon_l$, $X_{Nm}$, $V_{lm}$, and $\omega_{Nm}$ are determined by the orbital configuration and stellar spin.  

The total stellar response is determined by summing over all modes $\alpha$ for each orbital harmonic $N$. Details of this summation over oscillation modes are described in \citet{Fuller+17}, and involves summing over both negative and positive frequencies $\omega_\alpha$ for both negative and positive $m$.

\subsection{Stellar Models}

To determine $\omega_\alpha$ $Q_\alpha$, $L_\alpha$, and $\gamma_\alpha$ for the calculation of TEO amplitudes, we generated stellar models using \texttt{MESA} with stellar parameters close to those determined through binary modeling described in Section~\ref{sec:model}. Using the \texttt{GYRE} oscillation code \citep{Townsend+13, Townsend+18}, we calculated the $\ell=2$ non-adiabatic stellar oscillation modes of our \texttt{MESA} stellar models, and then computed $Q_\alpha$ and $L_\alpha$ as described in \cite{Fuller+17}. Since $L_\alpha$ is sensitive to the temperature perturbations due to a stellar mode $\alpha$ near the stellar photosphere, non-adiabatic modes are of interest to ensure accurate estimates of $L_\alpha$.

\subsection{TEO Amplitude}

The luminosity fluctuation amplitude $A_N$ is extremely sensitive to the detuning factor $\omega_{Nm}/\sqrt{(\omega_\alpha - \omega_{Nm})^2 + \gamma_\alpha^2}$ in Equation~\ref{eq:A_N}, with very small changes to the stellar model (and thus $\omega_\alpha$) causing significant changes to $A_N$. Thus, we follow \citet{Fuller+17}'s statistical approach where the mode amplitudes are predicted as a function of frequency. 

The median luminosity fluctuation is
\begin{equation}
    A_{N, \text{med}} \simeq \left|4 \mathcal{L}_N \frac{\omega_{Nm}}{\Delta{\omega_\alpha}} \right|,
\end{equation}
where $\Delta \omega_\alpha$ is the frequency spacing between successive $\ell=2$ g modes, and $\mathcal{L}_N = \epsilon_l V_{lm} X_{Nm} |Q_\alpha L_\alpha|$, both of which are evaluated for the mode most resonant with the forcing frequency $N \Omega$. We expect $95\%$ of TEOs exist between amplitudes of
\begin{equation}
    A_{N, \text{lower}} \simeq \left|2.05 \mathcal{L}_N \frac{\omega_{Nm}}{\Delta{\omega_\alpha}} \right|,
\end{equation}
and 
\begin{equation}
    A_{N, \text{upper}} \simeq \left|80 \mathcal{L}_N \frac{\omega_{Nm}}{\Delta{\omega_\alpha}} \right|,
\end{equation}
Additionally, the maximum theoretical amplitude of a TEO is
\begin{equation}
    A_{N, \text{theoretical max}} \simeq \left|\mathcal{L}_N \frac{\omega_{Nm}}{\gamma_\alpha} \right|.
\end{equation}

\subsection{KIC 6117415}

For our \texttt{MESA} model, we adopted our fitted binary model radius of $1.462~\text{R}_\odot$ and a mass of $m_1 = 1.3~\text{M}_\odot$ (within $11\%$ of the mass derived from binary model fitting). This slightly lower mass was chosen from the model that matches the inferred position of the primary on the Hertzsprung-Russell diagram shown in Figure~\ref{fig:6117HR}. We adopted the SpecMatch metallicity of $[\textrm{Fe/H}] = -0.1$. In our \texttt{GYRE} model, we used a rotation rate of $3.7~$days, determined using the best fit binary model primary radius and inclination and $v\sin{i} = 19.65~$km/s from SpecMatch. 

\begin{figure}
	\includegraphics[width=\columnwidth]{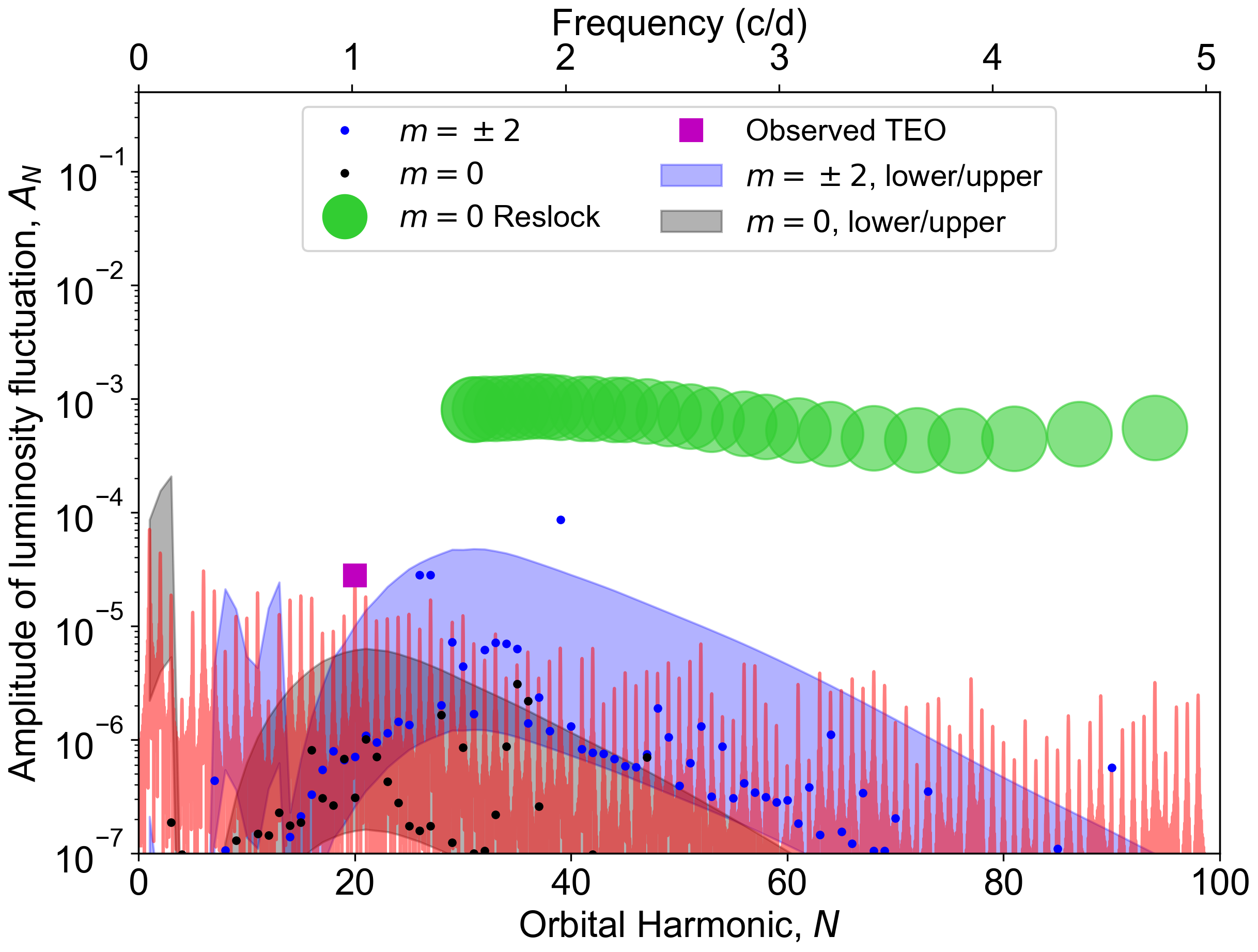}
	\caption{\textbf{TEO amplitudes as a function of frequency for KIC 6117415.} Blue represents modes with $|m| = 2$, and black represents modes with $m=0$. Dots denote representative $A_N$ values from our tidal models, and shaded regions mark where we expect $95\%$ of TEOs to occur. Only the contribution of the primary star is shown, as the slightly smaller secondary is expected to exhibit lower TEO amplitudes. The observed power spectrum of the system is shown in red. The regularly spaced peaks with $A_N \sim 10^{-5}$ occur at integer multiples of the orbital frequency and are likely remnants of our imperfect binary model. The best candidate TEO at $N=20$ is marked with a magenta square. The TEO amplitudes predicted by resonance locking of $m=0$ modes are marked in light green. Resonance locking of $m=2$ modes occurs at higher frequencies than those shown here.}
	\label{fig:6117A_N}
\end{figure}

Figure~\ref{fig:6117A_N} presents the predicted TEO amplitudes for KIC 6117415 as a function of frequency. As expected, the majority of the modeled luminosity fluctuation amplitudes exist within the $95\%$ expectation region. The possible observed TEO at $N=20$ lies slightly above the $95\%$ expectation region, suggesting that it may be explained by chance resonance between a stellar oscillation mode and the tidal forcing frequency. However, our model over-predicts the amplitude of modeled TEOs around $N=40$, predicting a few observable modes with amplitude greater than $10^{-5}$, whereas none are clearly detected. This may be due to model parameters such as stellar radius or temperature that are slightly too large. A smaller radius would imply a smaller value of the tidal forcing amplitude $\epsilon_l$, which scales as $R^3$. A lower temperature would also imply smaller amplitude TEOs, because the star would have a deeper convective envelope such that the surface luminosity fluctuations $L_\alpha$ would be smaller for the star's gravity modes.

\subsection{KIC 11494130}

We used our fitted binary model radius of $1.59~\text{R}_\odot$ and a mass of $m_1 = 1.40~\text{M}_\odot$ (within $4\%$ of the mass derived from binary model fitting, see Figure~\ref{fig:114HR}), and TLS temperature of $6500$~K to select our model. We used a rotation rate of $11.1~$days, determined using the best fit binary model primary radius and inclination and $v\sin{i} = 7.1~$km/s from TLS spectroscopy. 

\begin{figure}
	\includegraphics[width=\columnwidth]{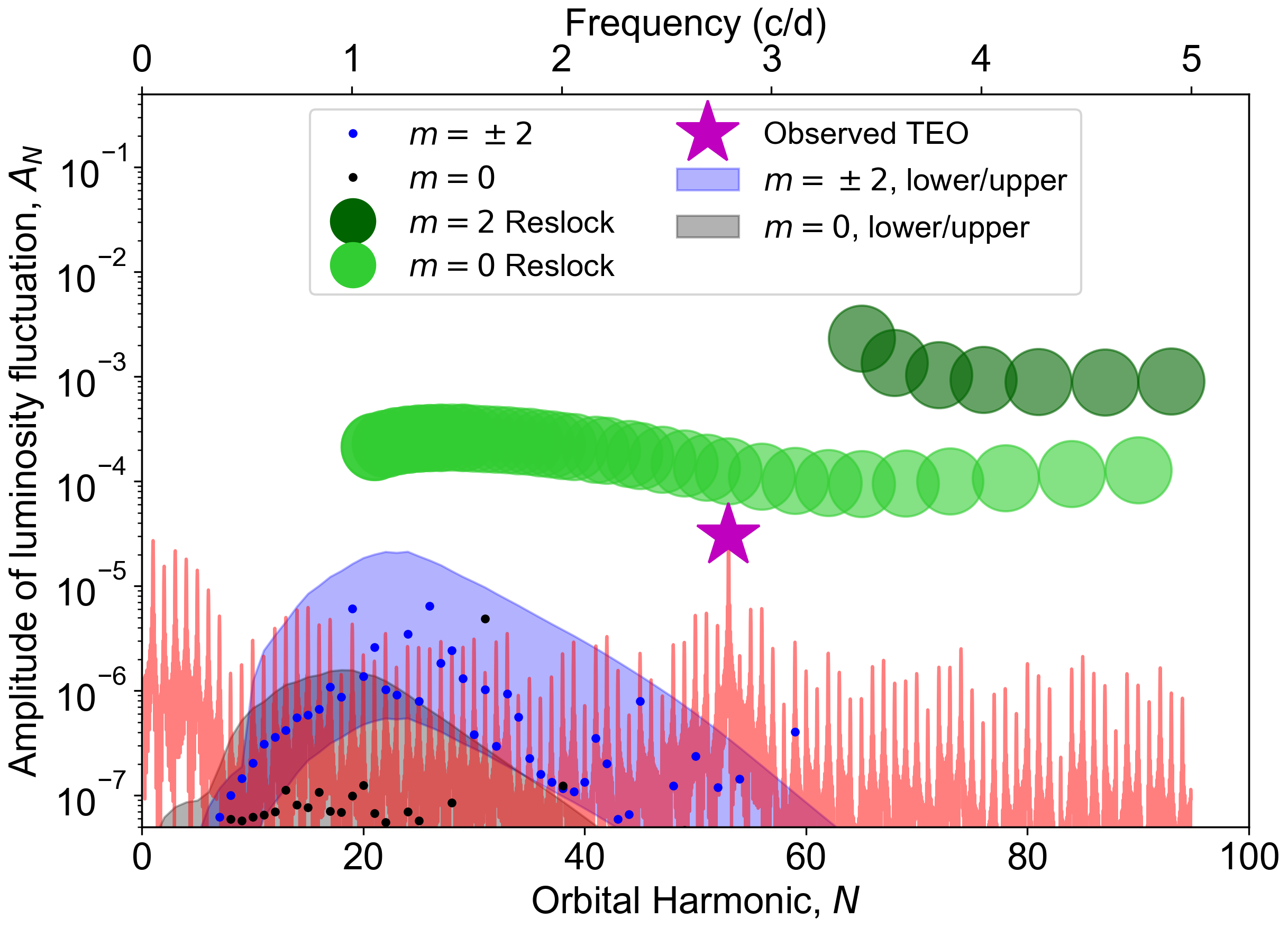}
	\caption{\textbf{TEO amplitudes as a function of frequency for KIC 11494130.} The structure here follows that in Figure~\ref{fig:6117A_N}. The dark green circles mark the luminosity fluctuation amplitudes predicted by resonance locking of $m=2$ modes. The magenta star highlights the frequency and luminosity fluctuation amplitude of the prominent TEO at $N=53$. The prominent TEO lies far above the expectation for a chance resonance, but close to the prediction for an $m=0$ resonantly locked mode.}
	\label{fig:114A_N}
\end{figure}

Figure~\ref{fig:114A_N} compares the predicted TEO amplitudes with periodograms of KIC 11494130. The observed high-amplitude TEO at $N=53$ is much larger than that predicted by our models at this frequency, assuming chance resonances between tidal forcing and stellar oscillation modes. The models predict that the largest amplitude TEOs should occur in the range $N=20-40$, and these TEOs are expected to lie near or below the detection threshold, consistent with the absence of any prominent TEOs in this frequency range. A chance resonance fails to account for the high amplitude TEO in KIC 11494130, so in Section \ref{sec:reslock} we explore the possibility of resonance locking \citep{Witte+99, Witte+01}.

\subsection{KIC 5790807}

To select our model, we used a temperature of $6466$~K from SpecMatch HIRES, a rotation rate of $9.3~$days (determined using $v\sin{i} = 14.56~$km/s from SpecMatch), a radius of $2.68~\text{R}_\odot$ from our binary fit, and a mass of $1.7~\text{M}_\odot$. This mass is within $3\%$ of the fitted binary model mass (see Figure~\ref{fig:57HR}). 

In KIC 5790807, the observed TEOs at $N=\ 48 \ \text{and } 107$ lie very close to or within the $95\%$ expectation region, at frequencies near those predicted by our model. Therefore, the observed TEOs in KIC 5790807 are consistent with the predictions of chance resonances. We again slightly overpredict the amplitudes of TEOs in the range $N=70-130$, as the models predict several TEOs with amplitudes above $10^{-5}$, but only two are clearly observed. Like KIC 6117415, this discrepancy is likely due to imperfect modeled parameters, and may result from a model radius or temperature that is slightly too large.

\begin{figure}
	\includegraphics[width=\columnwidth]{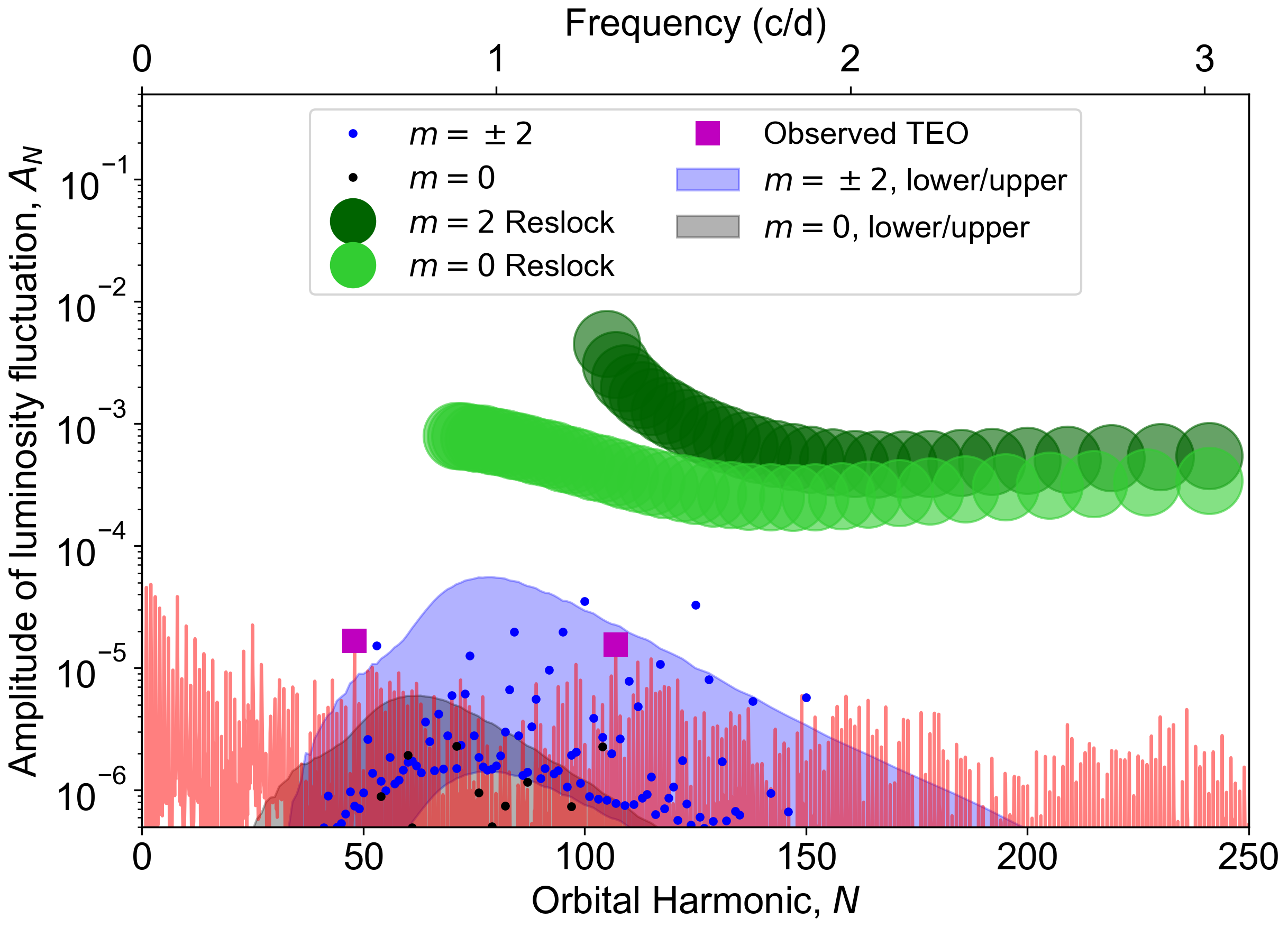}
	\caption{\textbf{TEO amplitudes as a function of frequency for KIC 5790807.} The structure here follows that in Figure~\ref{fig:114A_N}. The magenta sqaures at $N=48$ and $N=107$ are likely TEOs.}
	\label{fig:57A_N}
\end{figure}

\subsection{Resonance Locking}\label{sec:reslock}


When resonance locking occurs, a resonantly excited stellar oscillation mode causes the orbit to evolve at a rate such that the mode remains resonant, even as its frequency evolves due to the evolution of its host star. Therefore, the resonance is preserved and may be stable for extended periods of time \citep{Burkart+14}. In heartbeat stars, the g modes typically increase in frequency as the star ages due to the increasing Brunt-V\"ais\"al\"a frequency within the star. Resonantly locked modes thus produce tidal dissipation that causes the orbital frequency to increase at the rate necessary to maintain the resonance. 

To investigate whether resonance locking can explain the prominent TEO in KIC 11494130, we follow the theory detailed in \citet{Fuller+17}, where the resonance locking mode luminosity fluctuations are given by:
\begin{equation}
    A_{N, \text{ResLock}} = \left[\frac{c_\alpha}{\gamma_\alpha t_\alpha} \right]^{1/2} V_{lm} L_\alpha .
\end{equation}
In this expression, $V_{lm}$, $L_\alpha$ and $\gamma_\alpha$ are, as before (see Section~\ref{sec:tidalmodel}), the viewing angle correction factor, the luminosity fluctuation and damping rate of the resonantly locked mode. $c_\alpha$ is a dimensionless factor (with typical value of $\sim \! 10^{-2}$), as defined and derived in \citet{Fuller+17}. $t_\alpha$ is the mode evolution time-scale due to stellar evolution, and is defined as $t_\alpha = \sigma_\alpha / \dot{\sigma_\alpha}$ where $\sigma_\alpha = \omega_\alpha + m\Omega_s$ is the stellar mode frequency in the inertial frame, and $\Omega_s$ is the rotational frequency. To determine $\dot{\sigma_\alpha}$, we create \texttt{MESA} stellar models slightly younger and older than our model for KIC 11494130, and compute their stellar oscillation modes using \texttt{GYRE}. We then compute $t_\alpha = \Delta t (\sigma_\alpha/\Delta \sigma_\alpha)$, where $\Delta t$ is the difference in age between the models, and $\Delta \sigma_\alpha$ is the difference in frequency for modes with the same radial order $n$. We find typical values of $t_\alpha \sim 4~$Gyr for $m=0$ g modes in our model of KIC 11494130.

As shown in Figure~\ref{fig:114A_N}, resonance locking of $m=0$ modes produces luminosity fluctuations with amplitudes higher than but nonetheless comparable to the prominent TEO at $N=53$ in KIC 11494130. This discrepancy may be due to inaccurate fitting of stellar properties. Specifically, if the star actually has a slightly smaller radius (and is hence less evolved), we expect slower evolution, larger values of $t_\alpha$, and hence smaller resonance locking amplitudes. Additionally, a slightly lower temperature would lead to smaller values of $L_\alpha$ and could also resolve the discrepancy. Much larger changes to the models are needed to explain the mode via a chance resonance. We therefore conclude that resonance locking is likely to be occurring in KIC 11494130.

For consistency, we perform similar resonance locking calculations for KIC 6117415 and KIC 5790807. As shown in Figures~\ref{fig:6117A_N} and \ref{fig:57A_N}, the luminosity fluctuation amplitudes predicted by resonance locking are higher than those predicted by chance resonances. Additionally, a resonantly locked mode would have an amplitude of nearly $10^{-3}$ in these systems, much larger than our detectability threshold of about $5 \times 10^{-6}$, so any TEOs due to resonance locking would have been detected in the light curve. Therefore, since these systems do not exhibit any high amplitude TEOs, we can conclude that resonance locking is very unlikely to be occurring in KIC 6117415 or KIC 5790807.

\section{Discussion and Conclusions} 
\label{sec:discuss}

We characterized three \textit{Kepler} heartbeat stars (KIC 6117415, KIC 11494130, and KIC 5790807) using \textit{Kepler} photometric data and spectroscopic radial velocity measurements from Keck/HIRES, the TLS observatory, \citet{Smullen+15}, and \citealt{Shporer+16}. To do this, we created binary models of the heartbeat stars to simultaneously fit both the light curves and radial velocity measurements. We obtain reasonably good fits for the systems using the binary modeling software \texttt{ellc} \citep{Maxted+16} and \texttt{emcee} \citep{Foreman-Mackey+13}, a Markov chain Monte Carlo (MCMC) implementation. In our best MCMC fits, all parameters converged to approximately Gaussian distributions about reasonable parameters for main sequence heartbeat star systems. Our results are also largely consistent with temperature and luminosity information derived by \citet{Berger+18} from Gaia data and SpecMatch-Emp \citep{yee+17} fits to HIRES data. Best fit parameters for each system are provided in Tables \ref{table:6117mcmc}-\ref{table:57mcmc}.

KIC 6117415 is the only double-lined system we studied, and it is also eclipsing, so the light curve and radial velocity fits for KIC 6117415 were relatively easily obtained. Though KIC 5790807 is a single-line system, its eclipsing nature provides constraints on inclination and thus helps reduce the size of the parameter space, so MCMC convergence is obtained without great difficulty.
KIC 11494130 presents challenges during MCMC fitting due to the lack of an eclipse to constrain inclination, the single-lined nature of the system, and the presence of tidally excited oscillations (TEOs). By introducing constraints on the secondary star parameter priors motivated by physical understanding of main sequence stellar properties and mass-luminosity relations, we nonetheless obtain a good fit.

We investigated whether TEOs in these heartbeat stars are likely due to chance resonances with stellar oscillation modes, or the result of resonance locking between a tidal forcing and a stellar oscillation mode. We first identified TEOs in the heartbeat stars by performing frequency analysis on the residuals of our light curve models. In KIC 6117415, we identified a possible TEO at the $N=20$ orbital harmonic but did not identify any prominent TEOs. Similarly, in KIC 5790807, we did not observe any prominent TEOs but identify likely TEOs at $N=48 \ \text{and } 107$. In KIC 11494130, we identify a single prominent TEO at the $N = 53$ orbital harmonic whose amplitude is much larger than any other TEOs in this system.

To characterize the TEOs, we followed \citet{Fuller+17} and created stellar and tidal models of the heartbeat stars using the \texttt{MESA} stellar evolution code and the \texttt{GYRE} oscillation code. We then estimated the amplitude of TEOs through a statistical approach following \citet{Fuller+17}, which predicts the most probable TEO amplitudes as a function of frequency. In KIC 6117415 and KIC 5790807, the predicted TEO amplitudes and frequencies were similar to those measured, so the observed TEOs can likely be explained by chance resonances. In fact, our models slightly over-predicted the amplitudes of TEOs in these systems, likely because our inferred stellar radius or temperature was slightly too large. 

In KIC 11494130, the prominent TEO at the $N=53$ orbital harmonic has an amplitude far above the $95\%$ confidence range predicted by the tidal theory for chance resonances, which are unlikely to be responsible for the prominent TEO. We explored whether resonance locking \citep{Fuller+17} can explain the prominent TEO in KIC 11494130, finding that the luminosity fluctuation amplitude predicted by resonance locking of an $m=0$ g mode can likely explain the prominent TEO in KIC 11494130. Therefore, we conclude that resonance locking is likely to be operating in KIC 11494130. However, for KIC 6117415 and KIC 5790807, a resonantly locked TEO would have an amplitude far larger than the observed TEOs, so it appears resonance locking is not operating in those systems.

It remains unclear why resonance locking appears to be active in some heartbeat stars (e.g., KIC 11494130; KIC 8164262, \citealt{Fuller&Hambleton+17}; and perhaps in KOI-54, \citealt{Fuller+12a}), but not in others. For several systems (e.g., KIC 6117415 and 5790807; KIC 3230227, \citealt{Guo+17}; KIC 4142768, \citealt{Guo+19}) chance resonances can explain the observed TEOs, and resonance locking can be excluded by the lack of high-amplitude TEOs sufficient to induce a resonance lock. It is possible that resonance locking is a transient phenomenon, only occurring some fraction of the time because resonance locks are periodically broken due to resonance crossings with other modes. It also remains possible that resonance locking does not occur in many heartbeat stars because non-linear mode coupling \citep{Weinberg+12} truncates the resonant peaks, though we disfavor this possibility because much larger amplitude g modes are frequently observed in $\gamma$-Doradus and SPB pulsators. Unfortunately, it is difficult to model TEOs for a large population of heartbeat stars because of the large amount of data (space-based photometry and spectroscopic RVs) and modeling (light-curve and TEO analysis) needed for each system. Nonetheless, future TEO modeling may reveal trends in the systems exhibiting tidal resonance locking, allowing us to better understand its impact on binary stellar evolution.

\section*{Acknowledgments}

S.J.C would like to thank Susan Mullally for aiding in light curve processing, as well as Kevin Burdge and Erik Petigura for assistance with light curve modelling. This research is funded in part by a Heising-Simons Foundation 2018 Scialog grant (\#2018-1036), an Innovator Grant from The Rose Hills Foundation, and the Sloan Foundation through grant FG-2018-10515. KH and JF acknowledge support through a NASA ADAP grant (NNX17AF02G). The authors wish to recognize and acknowledge the very significant cultural role and reverence that the summit of Mauna Kea has always had within the indigenous Hawaiian community. We are most fortunate to have the opportunity to conduct observations from this mountain.

\appendix{}\label{appendix}

\begin{figure*}
	\includegraphics[width=\linewidth]{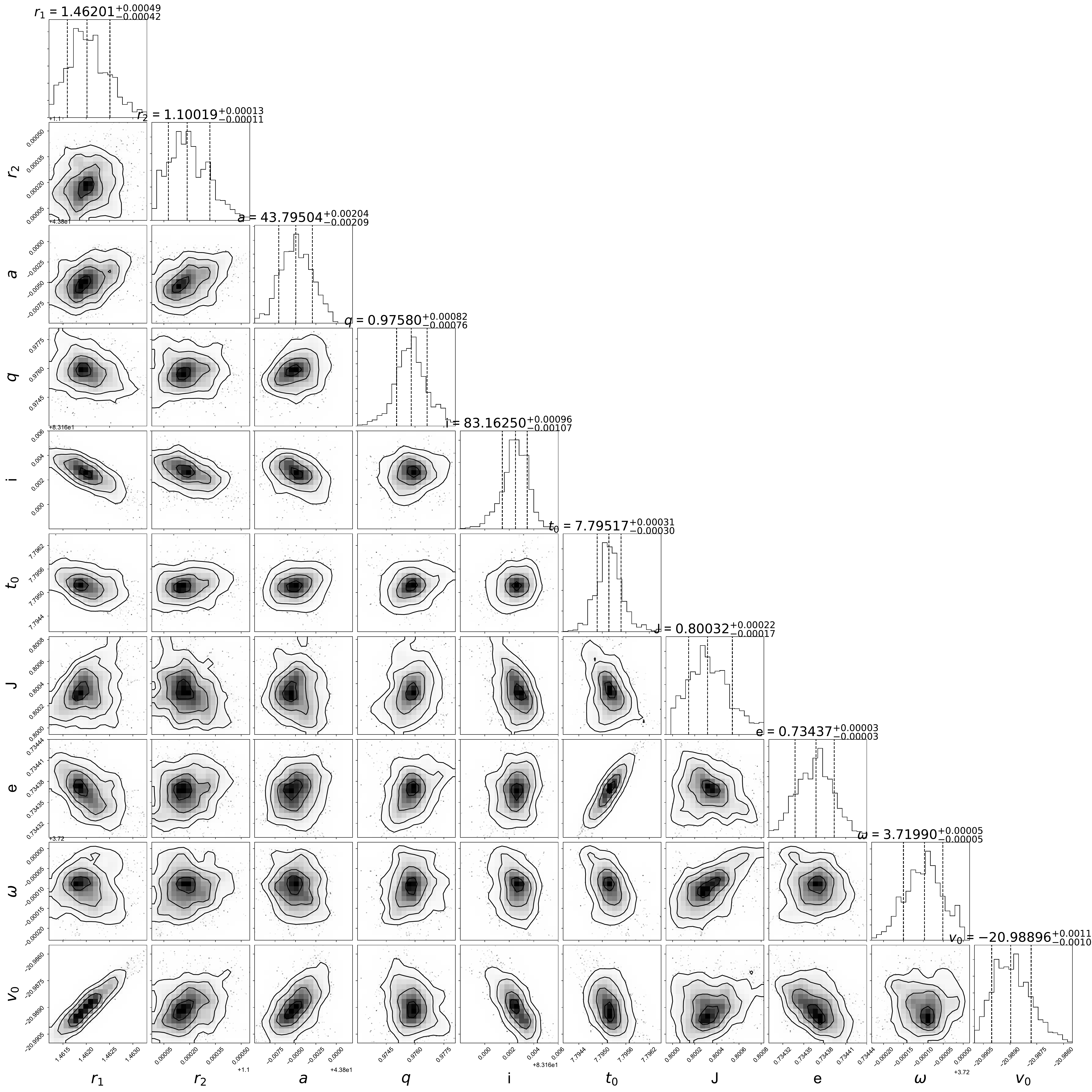}
	\caption{\textbf{Corner plot with posterior distributions of the best fit model parameters to KIC 6117415.} $r_1$ is the primary radius, $r_2$ is the secondary radius, $a$ is the semi major axis, $q$ is the mass ratio (secondary mass divided by primary mass), $i$ is inclination, $e$ is eccentricity, $\omega$ is the argument of periastron, $J$ is the surface brightness ratio (secondary divided by primary), $t_0$ is the time of mid-eclipse, and $v_0$ is the radial velocity offset. The best fit (mean) values and fitting uncertainties ($1\sigma$) for each parameter is given at the top of the column.
	A $2\sigma$ Gaussian image smoothing filter was applied to the images. Diagonal (from top left to bottom right): histograms showing the probability distributions of parameters, with dashed lines at the $0.16$ and $0.84$ quantiles. As in Table~\ref{table:6117mcmc}, these uncertainties are underestimated (see text).}
	\label{fig:6117corner}
\end{figure*}

\begin{figure*}
	\includegraphics[width=\linewidth]{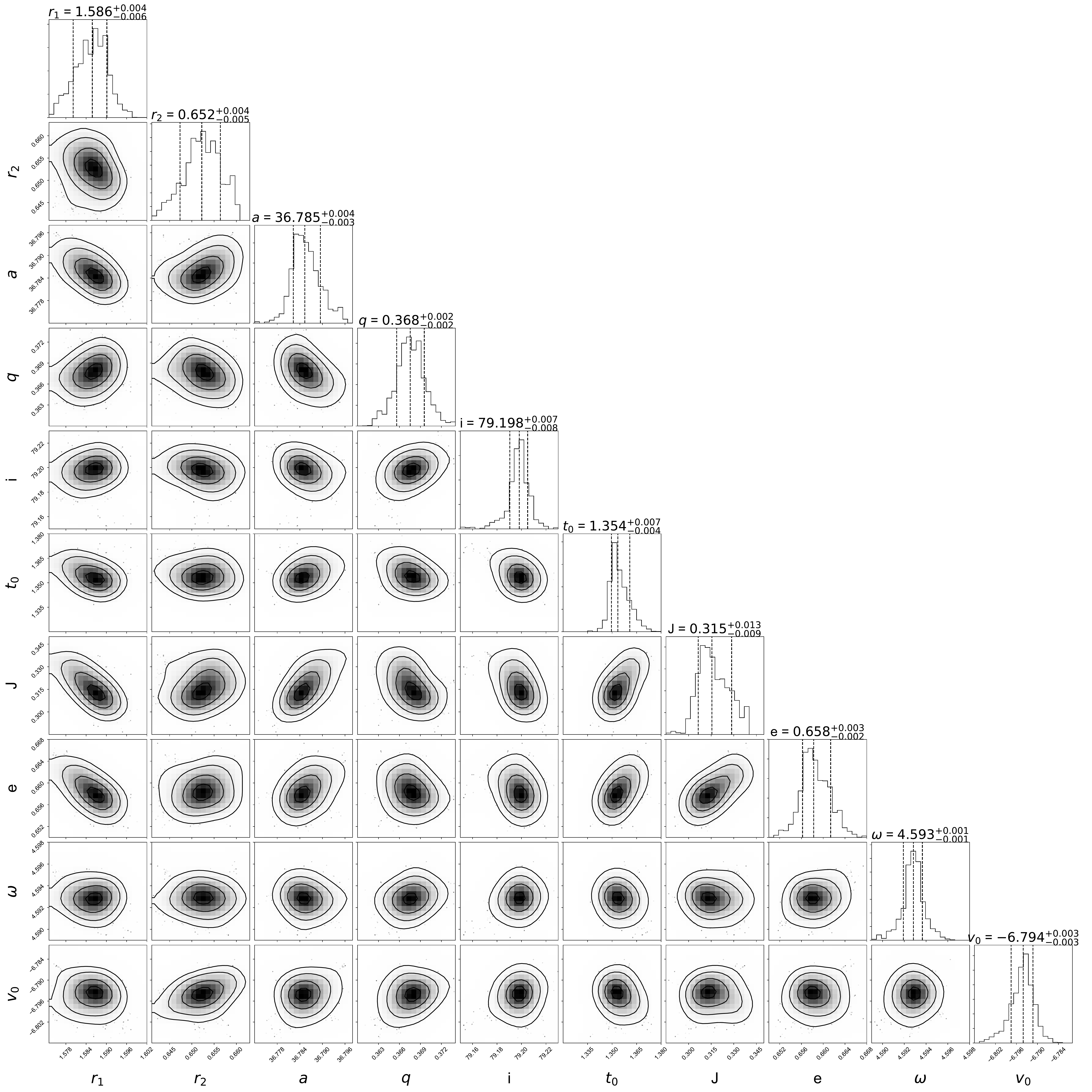}
	\caption{\textbf{Posterior distributions of the best fit model parameters to KIC 11494130.} This plot is identical in structure and content type as Figure~\ref{fig:6117corner}.}
	\label{fig:114corner}
\end{figure*}

\begin{figure*}
	\includegraphics[width=\linewidth]{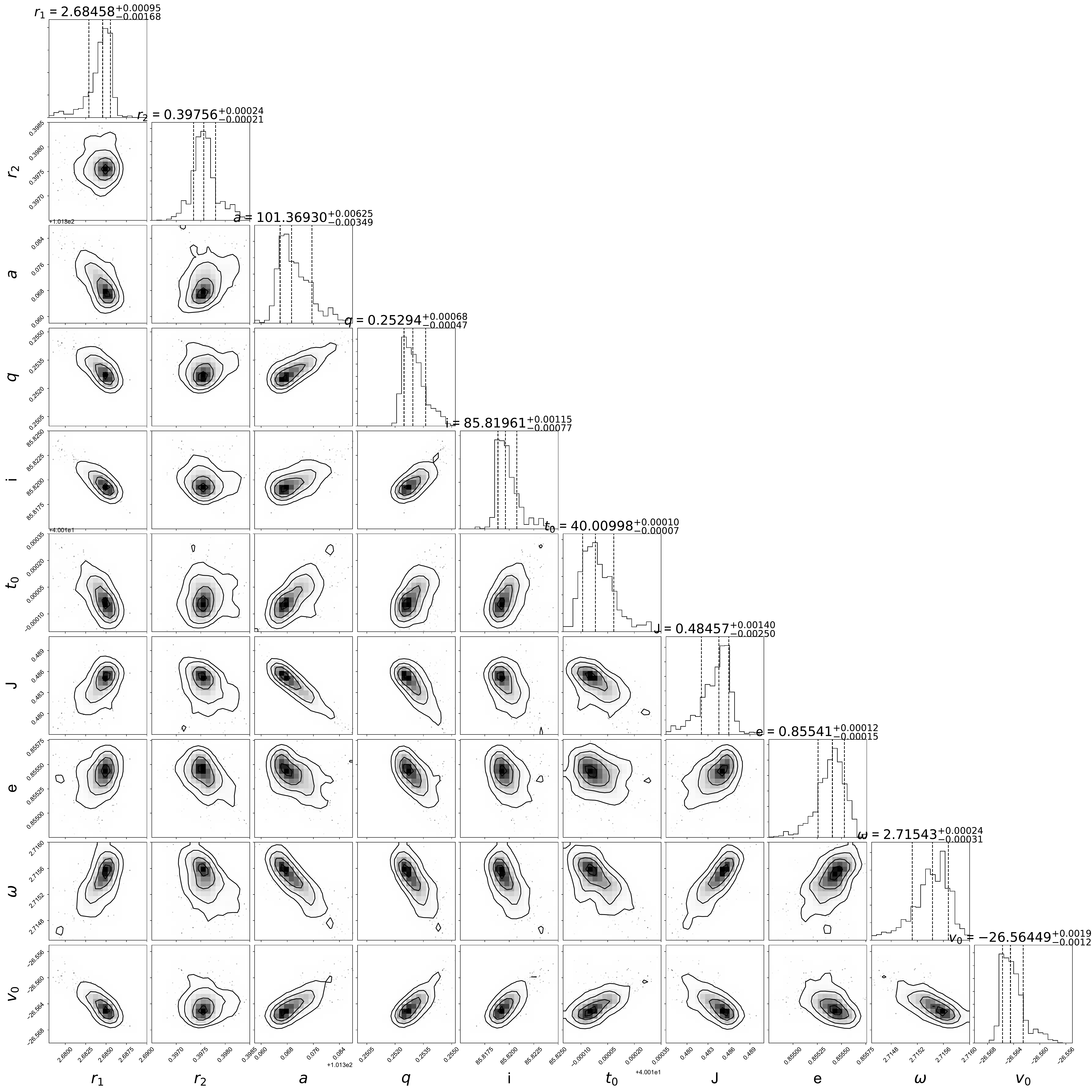}
	\caption{\textbf{Posterior distributions of the best fit model parameters to KIC 579080.} This plot is identical in structure and content type as Figures~\ref{fig:6117corner} and \ref{fig:114corner}.}
	\label{fig:57corner}
\end{figure*}

We show radial velocity measurements for KIC 6117415 and KIC 11494130 in Tables~\ref{table:6117rvs} and \ref{table:114rvs}. We also show the MCMC fit posteriors in Figures~\ref{fig:6117corner}, \ref{fig:114corner}, and \ref{fig:57corner}. For all systems, all fitted parameters converged well with approximately Gaussian posteriors.  

\begin{table*}[h]
\centering
\begin{tabular}{ccccc}
\hline
 & \multicolumn{2}{c}{Star 1} & \multicolumn{2}{c}{Star 2} \\
Relative Phase & Radial Velocity & Uncertainty & Radial Velocity  & Uncertainty \\
\hline
 -0.46 &   -7.48 & 0.21 & -33.20 & 0.12 \\
 -0.33 &  -13.43 & 0.10 & -27.56 & 0.15 \\
 -0.14 &  -49.09 & 0.12 &   8.75 & 0.24 \\
 -0.10 &  -70.62 & 0.16 &  31.35 & 0.26 \\
 -0.02 & -149.04 & 0.15 & 110.30 & 0.27 \\
 -0.02 & -146.69 & 0.14 & 108.80 & 0.27 \\
  0.07 &   -1.16 & 0.13 & -44.87 & 0.18 \\
  0.11 &    4.85 & 0.12 & -49.87 & 0.23 \\
  0.17 &    5.99 & 0.15 & -51.70 & 0.27 \\
  0.34 &    2.66 & 0.14 & -44.62 & 0.21 \\
  0.39 &    0.99 & 0.13 & -41.69 & 0.17 \\
\hline
\end{tabular}
\caption{Radial velocities for the primary star in KIC 6117415. The data shown here were obtained following the procedure outlined in Section~\ref{sec:keck}. 1$\sigma$ uncertainties are shown, and all radial velocities and uncertainties are in units of km/s. The quoted relative phase is zero at periastron.}
\label{table:6117rvs}
\end{table*}

\begin{table*}[h]
\centering
\begin{tabular}{cccccc}
\hline
Relative Phase & Radial Velocity & Uncertainty & Relative Phase & Radial Velocity & Uncertainty \\
\hline
 -0.49 &  -1.80 & 4.30 & 0.18 & 16.60 & 4.20 \\
 -0.32 & -18.60 & 4.30 & 0.18 & 16.14 & 0.10 \\
 -0.24 & -22.70 & 4.20 & 0.18 & 12.50 & 4.20 \\
 -0.24 & -20.00 & 4.20 & 0.20 & 11.90 & 4.20 \\
 -0.14 & -30.50 & 4.20 & 0.24 & 11.70 & 4.20 \\
 -0.13 & -33.70 & 4.20 & 0.25 & 10.58 & 0.10 \\
 -0.13 & -35.32 & 0.10 & 0.30 &  6.75 & 0.10 \\
 -0.12 & -39.40 & 4.20 & 0.30 &  6.63 & 0.10 \\
 -0.09 & -36.60 & 4.20 & 0.35 &  3.70 & 4.20 \\
 -0.08 & -39.40 & 4.20 & 0.35 &  3.20 & 0.10 \\
 -0.07 & -42.40 & 0.10 & 0.40 &  0.99 & 0.10 \\
 -0.07 & -40.10 & 4.20 & 0.41 &  1.90 & 4.20 \\
 -0.03 & -35.10 & 4.20 & 0.41 & -0.65 & 0.10 \\
 -0.02 & -29.70 & 4.20 & 0.45 & -1.83 & 0.10 \\
  0.04 &  21.43 & 0.10 & 0.00 &  0.00 & 0.00 \\
\hline
\end{tabular}
\caption{Radial velocities for KIC 11494130. The format here follows Table~\ref{table:6117rvs}. The data shown here are derived from TLS spectroscopy data as described in Section~\ref{sec:TLS} and from \citet{Smullen+15}.}
\label{table:114rvs}
\end{table*}

\clearpage
\bibliography{heartbeat}{}
\bibliographystyle{aasjournal}

\end{document}